*Article*

# PCA-MRM model to forecast TEC at middle latitudes

Anna L. Morozova [1,*], Teresa Barata [2] and Tatiana Barlyaeva [3]

1. Instituto de Astrofísica e Ciências do Espaço, Univ Coimbra, OGAUC, Coimbra, Portugal; annamorozovauc@gmail.com
2. Instituto de Astrofísica e Ciências do Espaço, Univ Coimbra, OGAUC, Coimbra, Portugal; mtbarata@gmail.com
3. Instituto de Astrofísica e Ciências do Espaço, Univ Coimbra, OGAUC, Coimbra, Portugal; tvbarlyaeva@gmail.com
* Correspondence: annamorozovauc@gmail.com

**Abstract:** The total electron content (TEC) over the Iberian Peninsula was modelled using PCA-MRM models based on decomposition of the observed TEC series using the principal component analysis (PCA) and reconstruction of the daily modes' amplitudes by a multiple linear regression model (MRM) using space weather parameters as regressors. The following space weather parameters are used: proxies for the solar UV and XR fluxes, number of the solar flares of different classes, parameters of the solar wind and of the interplanetary magnetic field, and geomagnetic indices. Time lags of 1 and 2 days between the TEC and space weather parameters are used. The performance of the PCA-MRM model is tested using data for 2015, both geomagnetically quiet and disturbed periods. The model performs well for quiet days and days with solar flares but without geomagnetic disturbances. The MAE and RMSE metrics are of the order of 3-3.5 TECu for daytime and ~2 TECu for night-time. During geomagnetically disturbed periods the performance of the model deteriorates but only for daytime: MAE and RMSE are of the order of 4.5-5.8 TECu and can rise to ~13 TECu for the strongest geomagnetic storms.

**Keywords:** TEC model; middle latitudes; space weather; principal component analysis

## 1. Introduction

Space Weather (a group of phenomena observed in the near-Earth space and atmosphere that are related to variations of the solar activity) is one of the main drivers of ionospheric disturbances which, in turn, can drastically affect the quality of the signal between a ground-based device and the GNSS satellites. Forecasting and early warning for potentially dangerous space weather events are essential for the improvement of the quality of the GNSS-based services [1]. On the other hand, the GNSS receivers themselves are a reliable source of ionospheric data. Data from GNSS receivers can be utilized to estimate such widely used ionospheric parameters as the total electron content (TEC), scintillation indices (S4, $\sigma_\phi$) and ROTI (rate of TEC index).

The forecasts of the ionospheric parameters are based on the understanding of the ionosphere reaction to different forcings. Empirical (based on the observational data and their statistical analysis) models for ionospheric response to external forcings (e.g., solar flares and geomagnetic storms) are being developed by different research groups for their countries. The principal component analysis (PCA, also known as the empirical orthogonal functions (EOF) or the natural orthogonal components (NOC)) and different kinds of regression analyses are often used to model and forecast TEC variations using some space weather parameters as regressors or predictors.

To name a few, the authors of [2-3] proposed EOF-based models to forecast monthly median TEC in the form of regional ionosphere maps (RIMs) for China using the F10.7 solar index as predictor. Their models provided TEC forecasts with the standard deviation (SD) of 1-4 TECu (1 TECu = $10^{16}$ el/m$^2$), with lower SD during solar activity minima and higher SD for years of solar activity maxima.

In [4] a neural network (NN) based models were proposed that used F10.7 index, sunspot numbers (SSN) and a set of proxies for the solar UV flux (e.g., Mg II index) as TEC predictors. Their best model provided the root mean squared error RMSE = 3 TECu. They also showed a delayed response of TEC variations to the solar forcing: best results were obtained if SSN, F10.7, and solar UV proxies lagged backward relatively to TEC series by 1-2 days.

The authors of [5] used an EOF-based model to reconstruct global ionosphere maps (GIMs). They decomposed TEC observations into EOF functions that change with local time and dip latitude to represent the diurnal variation and spatial distribution of the original data and fitted their amplitude coefficients (that indicate the long-term temporal fluctuations) by F10.7 and geomagnetic indices Ap, AE and Dst. The model provided RMSE = 3-5% which for middle latitudes gives RMSE ≈ 2-3 TECu. In [6] PCA was also used to decompose GIM TEC variations (2007-2016) into several spatio-temporal modes fitting their amplitude coefficients by Ap and F10.7 indices. The ME (mean error) was in the range from -15 to 10 TECu and RMSE was in the range from below 5 to ~15 TECu for different years with worst errors observed during 2015, the year of strongest geomagnetic storms of the solar cycle 24.

The daily mean TEC (GIMs) was modelled in [7] using SSN and F10.7 as predictors. Their climatological model showed that the maximal differences between the observed and modelled TEC values are associated with geomagnetic storms (~3.2 ± 1.5 TECu). The authors of [8] proposed a GIM model to study effect of geomagnetic activity on the ionospheric conditions. They modelled the relative TEC (relative deviation from a 15-day median) using the Kp geomagnetic index as predictor (RMSE = 4.6 TECu).

Another EOF-based model by [9] was proposed to model TEC variations for geomagnetically quiet months with Ap and F10.7 indices as predictors of TEC variations: the mean average error MAE = 1.2-2.6 TECu. Storm-time TEC variations were modelled by [10] for mid-low latitudes using EOF and NN with F10.7 and A-type indices as predictors for TEC variations. The RMSE values were between 2 and 10 TECu for different storms occurred between 2000 and 2015 (with Dst ≤ -50 nT).

TEC variations at the European middle latitudes during quiet periods of 2015 were modelled by [11] using a number of models of different types. Their models that use space weather parameters as predictors give for the 40ºN latitude band RMSE = 3.08-3.82 TECu, ME = -0.4 - 0.1 TECu, median error = -0.1 - 0.3 TECu and the maximum error MaxE = 12-25 TECu. Another model that was tested on the data of 2015 was made by [12] for Balkan region. They used Kp and F10.7 indices as representatives of the solar and geomagnetic activity. They also showed delayed response of the ionosphere to geomagnetic storms (lag of 1-2 days for Kp). RMSE were in the range from 2.45 to 3.13 TECu and different between the night- (RMSE = 1.34-1.84 TECu) and daytime (RMSE = 4.5-5.5 TECu).

The single-station TEC (measured at a single location) was modelled by [13] using the F10.7 index as a proxy for the solar/geomagnetic activity (regression models) with RMSE = 3.22-4.46 TECu for different locations.

Almost all TEC models mentioned above, except, to some extent, those presented in [5-6], use additionally the information on the day of a year (DOY) and the hour to be able to model TEC daily and seasonal variations. The climatological models and the model built for geomagnetically quiet periods show, in general, better performance (lower RMSE and other errors) than models that use all or geomagnetically disturbed periods. There are no significant differences in the performance of the models developed using data from a single station and those using RIMs/GIMs. Most of the models described above are built using several years of TEC and space weather data. Some models incorporate lags of 1-2 days between the TEC response to space weather variations. Most often used space weather parameters are F10.7 and Mg II indices as proxies for the solar UV irradiance and A-type indices to account for the geomagnetic activity variations. Overall, RMSE are in the range from 2 to 15 TECu with higher RMSE values obtained for models that simulate TEC variations during geomagnetic storms.

Here we present the analysis of the performance of a new TEC model which is based both on the PCA and regression analyses. We use a different approach to model TEC daily and seasonal variations and used a short (~30 days) time interval to develop a model. Also, we use a large set of space weather parameters as predictors of TEC variations. Our model does not distinguish between geomagnetically active and quiet days.

The paper is organized as follows: Section 1 gives a short summary of the performance of models developed to simulate TEC variations using space weather parameters as predictors; Section 2 describe the data used to build and test our TEC model; Section 3 describe methods utilized in our model and metric we used to test its performance; Sections 4 and 5 describe the proposed model and its performance, respectively; Section 6 presents main conclusions.

**2. Data**

To build and test our model we used following data sets from different ground and space missions.

*2.1. Total electron content (TEC)*

The TEC series is obtained at Lisbon (Portugal) using a GNSS receiver with the SCINDA software [14-17] that has been active from the November 2014 to the July 2019 in the Lisbon airport (38.8ºN, 9.1ºW) in the frame of the ESA Small ARTES Apps project SWAIR (Space Weather and GNSS monitoring services for Air Navigation). The installed equipment was a NovAtel EURO4 receiver with a JAVAD Choke-Ring antenna. The data originally of 1 min time resolution were averaged to obtain the 1h series. In this work we used only data from January 1 to December 31 of 2015. The raw data were processed using a "SCINDA-Iono" toolbox for MATLAB [18] developed by our group and are available at [19]. This dataset is also described in [20] where TEC variations related to the solar flares and geomagnetic disturbances of 2015 were analysed.

The calibration procedure was not performed during the installation of this receiver. Therefore, we performed a calibration of the TEC records using the TEC data from the Royal Observatory of Belgium (ROB) as a reference. The Royal Observatory of Belgium (ROB) data base ([ftp://gnss.oma.be/gnss/products/IONEX/](ftp://gnss.oma.be/gnss/products/IONEX/)) provides the vertical TEC maps in the IONEX format on a grid of 0.5° x 0.5° with 15 min time resolution [21]. The vertical TEC is estimated in near real-time from the GPS data of the EUREF Permanent Network (EPN). To perform the calibration of our TEC series we used the ROB TEC data for the grid point most close to Lisbon (39ºN, 9ºW) using 1h mean ROB TEC series. The calibration was performed individually for each of the 12 months. The calibration coefficients were in the range from 1.74 to 2.34 depending on the month allowing to convert the relative TEC SCINDA data [19] to the TEC data in TECu. Since for Lisbon UT = LT, no time conversion was applied.

*2.2. Space weather data*

Preliminary tests of the PCA-MRM models as well as the results of [20] were used to define the set of space weather parameters (SWp) that are used as predictors for TEC variations. In particular, we prefer to use the *Mg II* index over the more often used F10.7 index [20]; also, the test models that used local $K_{COI}$ showed slightly better performance comparing to ones that used the global Kp index. The separation of the number of flares of different classes (C, M and N) was proposed to find out what is more important for a TEC model: the total number of flares, the number of the most abundant flares (C-class or less) or the number of moderate flares (M class); since in 2015 was only one short-living X-class flare the influence of the X flares on the model's performance was not studied.

Three types of SWp were used to forecast TEC variations:
1. Parameters characterising the solar UV and XR fluxes:
   a. a proxy for the solar UV irradiance: the Mg II composite series [22] based on the measurements of the emission core of the Mg II doublet (280 nm), hereafter *Mg II*;

b. a proxy for the solar XR irradiance: the measurements of the Solar EUV Experiment (SEE) for the NASA TIMED mission at the wavelength 0.5 nm, hereafter *XR*;
   c. the daily number of the solar flares of the classes up to C, hereafter *Number of C flares* or *C*;
   d. the daily number of the solar flares of the class M, hereafter *Number of M flares* or *M*;
   e. total daily number of the solar flares of any class, hereafter *Number of flares* or *N*;
2. Parameters characterising the interplanetary medium:
   a. scalar of the interplanetary magnetic field (IMF), *B* in nT;
   b. the X, Y and Z components of IMF in GSM frame, $B_X$, $B_Y$ and $B_Z$, respectively, all in nT;
   c. the solar wind flow pressure (*p* in nPa), proton density (*n* in n/cm$^3$) and plasma speed (*v* in km/sec).
3. Geomagnetic indices:
   a. the disturbance-time index *Dst*;
   b. the global *ap* index;
   c. the local K-index calculated from the horizontal component of the geomagnetic field measured at the Coimbra Magnetic Observatory, COI, Coimbra, Portugal, 40.2°N, 8.4°W, 99 m asl, $K_{COI}$;
   d. the auroral electrojet index *AE* characterizing the auroral activity in the polar regions.

The data on the solar wind properties were obtained from the OMNI data base. The information about the solar flares observed during the analysed time interval was obtained through the NOAA National Geophysical Data Center (NGDC). Only flares occurred during the local daytime were considered.

All SWp series used in the PCA-MRM model have 1d time resolution.

## 3. Methods

The PCA-MRM model is based on a combination of the principal component analysis (PCA) applied to the 1h TEC series and multiple regression models (MRM) to forecast amplitudes of TEC daily variations using SWp as regressors. The quality of the forecast of the TEC series was characterised by a set of standard metrics. These methods are described below.

*3.1. PCA*

The analysis of the TEC series was performed using the principal component analysis (PCA). The input data set is used to construct a covariance matrix and calculate corresponding eigenvalues and eigenvectors. The eigenvectors are used to calculate principal components (PC) and empirical orthogonal functions (EOF). The eigenvalues allow to estimate the explained variances of the extracted modes. PCs are orthogonal and conventionally non-dimensional. The full descriptions of the method can be found in (e.g.) [23-25].

The PCA input matrices were constructed in a way that each column contains 24 observations (means for every 1 h) for a specific day. Daily mean TEC values were removed from the series submitted to PCA. The number of the columns, *L*, is equal to the length of the studied interval in days. Thus, PCA allows us to obtain daily variations of different types as PCs and the amplitudes of those daily variations for each day of the studied time interval as corresponding EOFs. Only the 1st and 2nd PCA modes, PC1 and EOF1, and PC2 and EOF2, respectively, were used to reconstruct TEC variations [20].

*3.2. Regression models*

The linear multiple regression models (MRM) were constructed using TEC-related parameters, i.e., the daily mean TEC (1d mean TEC), and the EOF1 and EOF2 series, as dependent variable, and SWp as regressors. A "best subset" technique was chosen to ensure that only those regressors that are most influential for a particular TEC series were selected. The "best subsets" were estimated by minimizing a so-called adjusted squared coefficient of multiple determination ($R_{adj}^2$).

*3.3. Metrics for the forecast quality*

Similarities between the forecasted and observed TEC or other series were analyzed using the Pearson correlation coefficients, *r*. The significance of the correlation coefficients was estimated using the Monte Carlo approach with artificial series constructed by the "phase randomization procedure" [26]. The obtained statistical significance (*p value*) considers the probability of a random series to have the same or higher absolute value of r as in the case of a tested pair of the original series.

The quality of the forecast was also estimated using the following parameters: the root mean squared error *RMSE* (Eq. 1), the explained variance *ExpV* (Eq. 2), the coefficient of determination $R^2$ (Eq. 3), the mean absolute error *MAE* (Eq. 4), the mean error *ME* (Eq. 5), the maximum error *MaxE* (Eq. 6), the *range*, *mean* and *median* values of the differences between the forecasted and observed TEC values (*ΔTEC*), the percentage of the forecasted series with ΔTEC in certain limits. The forecasting quality of a model considered to be better if it had lower values of MAE and RMSE, higher values of r, $R^2$ and ExpV, and higher percentage of days with ΔTEC in a certain range.

$$RMSE = \sqrt{\Sigma(y_i - \hat{y}_i)^2/N}, \qquad (1)$$

$$ExpV = 1 - \sigma_{\Delta y}^2/\sigma_y^2, \qquad (2)$$

$$R^2 = 1 - \Sigma(y_i - \hat{y}_i)^2/\Sigma(y_i - \bar{y})^2, \qquad (3)$$

$$MAE = |\Delta TEC| = \Sigma|y_i - \hat{y}_i|/N, \qquad (4)$$

$$ME = mean\ \Delta TEC = \Sigma(y_i - \hat{y}_i)/N, \qquad (5)$$

$$MaxE = max(|y_i - \hat{y}_i|), \qquad (6)$$

where $y_i$ and $\hat{y}_i$ are the observed and modeled series, respectively; $\bar{y}$ and $\sigma_y$ are the mean and the standard deviation (SD) for $y_i$, respectively; $\sigma_{\Delta y}$ is SD for the ΔTEC = $(y_i - \hat{y}_i)$ series; and N is the length of the series.

## 4. PCA-MRM model for TEC forecasting

*4.1. General description of the PCA-MRM TEC model*

The TEC forecasting using a PCA-MRM model consists of three stages (Fig. 1). At the first stage the TEC data obtained for a certain time interval *L* (e.g., previous 20-40 days) are decomposed with PCA into several modes each one representing a specific type of daily variation (PCs). Only two first modes, which have highest variance fractions and are responsible for most of the TEC variability, are used for further analysis [20]. One of the advantages of the model is that there is no need for any assumption on the phase and amplitude or seasonal/regional features of TEC daily variations: the daily variations of correct shapes are extracted automatically by PCA from the input TEC data. The examples of PCs can be found in the Supplementary Material (SM), Fig. S1.

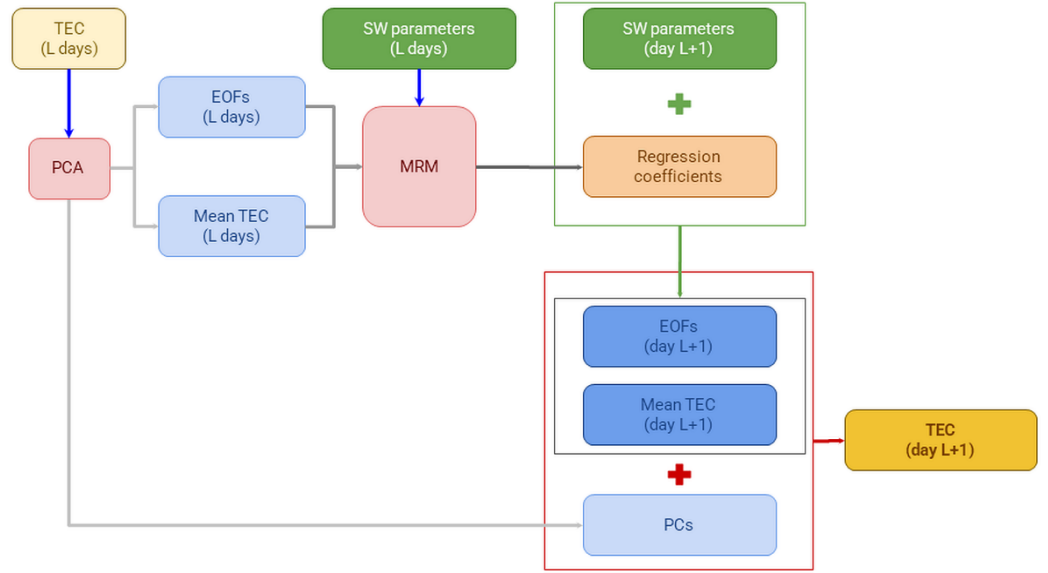

**Figure 1.** The PCA-MRM TEC model scheme.

The amplitude of each of these modes (EOFs) varies from day to day. During the second stage, these EOFs as well as the 1d mean TEC series are submitted to the regression analysis to construct MRMs that make correspondences between these TEC parameters and the variations of SWp selected as predictors. Every time a "best subset" of predictors is estimated to maximize the $R_{adj}^2$ of the resulting regression model. MRMs are constructed using a lag of 1 or 2 days between the variations of SWp and TEC (SWp lead).

As a result, the regression coefficients are generated allowing to use them at the final, third, stage to reconstruct (forecast) TEC for the following day, day $L$+1: we use correspondingly lagged SWp series as predictors to forecast the daily mean TEC, EOF1 and EOF2 values for that day and combining them with PC1 and PC2 to reconstruct (forecast) the 1h TEC series for the day $L$+1. No negative 1d mean TEC and EOF1 series were allowed: in case MRMs forecast negative values of 1d mean TEC or EOF1 they were multiplied by -1. The PCA-MRM models are denoted as PCA-MRM($L$##, lag#) where $L$## is the length of the input time interval in days and lag# is the lag between the TEC and space weather parameters in days.

The 1h TEC series forecasts were made separately for the MRM models with lags of 1 and 2 days, and the final forecast is constructed as the arithmetic mean of these forecasts, hereafter PCA-MRM($L$##, mean.lag1.2).

To build and validate our model, we used the TEC data observed between January 1 and December 31 of 2015 in Lisbon airport. The assessment of the PCA-MRM model forecasting quality is presented in Sec. 5.

*4.2. Length of the input dataset L*

The length $L$ of the input data series, both TEC and SWp, strongly affects the model performance. The shorter length may result in better representation of the TEC daily modes but will not be sufficient for the construction of reliable MRMs with so many regressors. On the other hand, larger $L$ will allow to constrain well the regression coefficients, but the TEC daily modes may be resolved with lower quality because of the seasonal changes of the TEC daily variation. We made tests for the PCA-MRM performance varying the $L$ parameter from 25 to 45 days comparing the observed and forecasted series of 1h TEC, 1d mean TEC and the daily maximum (1d max) TEC using metrics listed in Section 3.3 (some examples can be found in SM, Figs. S2-S3).

Overall, the lowest and the highest $L$ values performed badly, and the best results were obtained for $L$ in the interval from 28 to 33 days. Neither of the models has best performance with all metrics, however the models with $L$ equals to 31 or 32 days score better when all metrics are taken into consideration. Therefore, all further models were constructed using *L* = 31 and *L* = 32.

**5. Performance of the PCA-MRM TEC model**

The forecasting quality of the models were studied on the hourly (1h TEC series) and the daily (1d mean and 1d max TEC series) time scales, during quiet (no solar flares, no geomagnetic disturbances) days, days with solar flares and days with geomagnetic disturbances, during different months, and during different hours of a day.

*5.2. General performance*

Figures 2 and 3 show the observed (black lines) and forecasted using the PCA-MRM(L31, mean.lag1.2) model (red lines) 1 mean and 1d max TEC series, respectively. Figures 4 and 5 show examples of the forecast for the 1h TEC series made with the PCA-MRM(L31, mean.lag1.2) model for June and February of 2015, respectively, the months with the worst and the best, respectively, forecasts for the 1h TEC series. The plots for all months of 2015 (including those for February and June) can be found in SM, Figs. S4-S14. The results for the PCA-MRM(L32, mean.lag1.2) model are very similar and not shown.

A simple visual analysis shows that the biggest differences between the observed and forecasted TEC series are seen during the days of geomagnetic storms: most prominently this is seen for the March (Fig. S5) and June (Fig. 4 and Fig. S8) – the months of the strongest storms of the solar cycle 24.

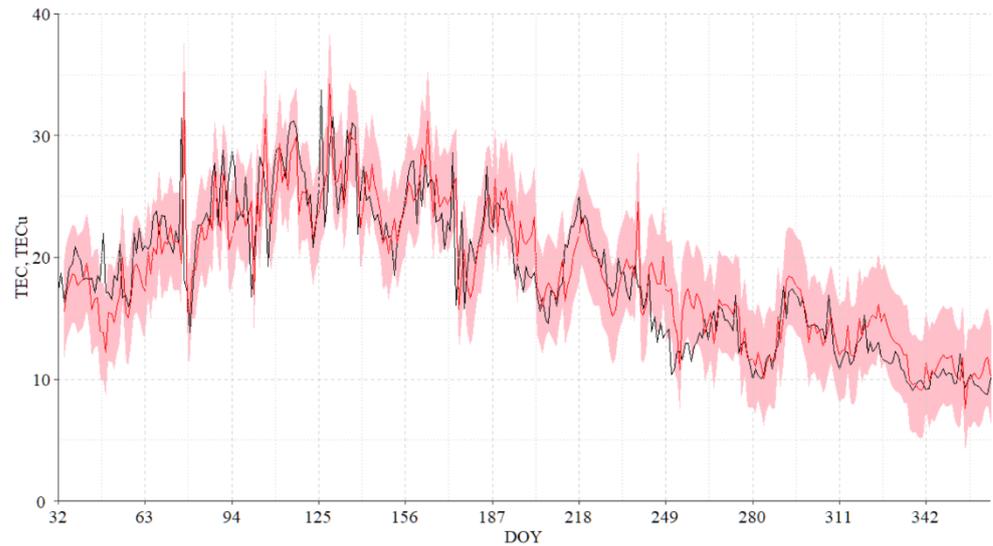

**Figure 2.** Observed (black) and forecasted using the PCA-MRM(L31, mean.lag1.2) model (red) daily mean TEC series; pink area shows 90% confidence interval.

Table 1 shows the mean and the range values for the 1h, 1d mean and 1d max observed and forecasted TEC series. As one can see, both the PCA-MRM(L31, mean.lag1.2) and PCA-MRM(L32, mean.lag1.2) models fit the observations similarly well but the performance of the PCA-MRM(L31, mean.lag1.2) model is slightly better. The scores (metrics values) of the two PCA-MRM models are shown in Tab. 2. Again, both models perform similarly well but the model with *L* = 31 days performs slightly better.

It also seems that the ability of the PCA-MRM models to forecast the 1d mean TEC values are better than for the 1h series: as shown in Tab. 2, the values of the mean and median ΔTEC, MaxE and RMSE values are about 1.5-2 times smaller for the 1d mean TEC series than for the 1h TEC series.

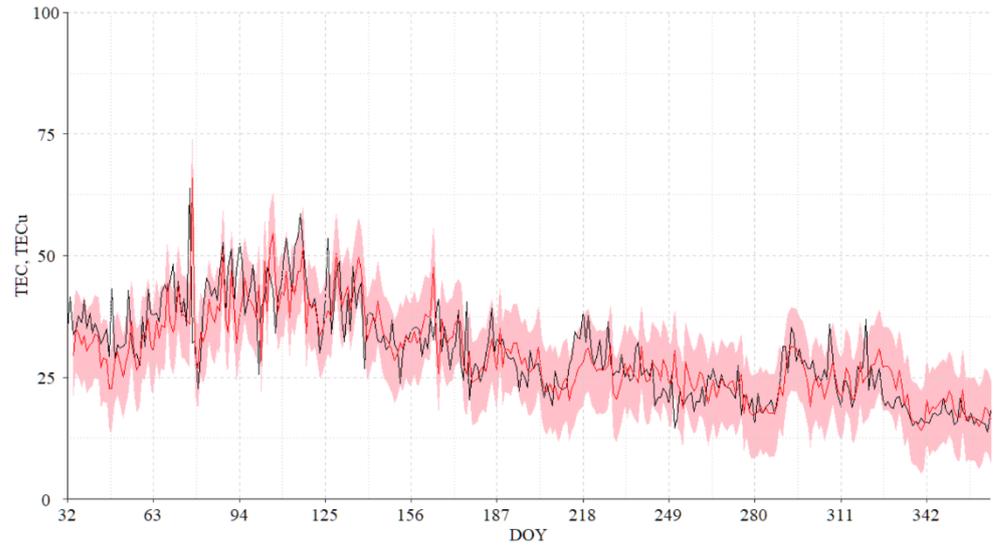

**Figure 3.** Same as Figure 2 but for the daily maximum TEC series.

**Table 1.** The mean values and the range (in TECu) for different TEC parameters for the observed and forecasted series: 1h TEC series, 1d mean and 1d maximum TEC series. Metrics of models that are most close to the observed ones are in bold.

| TEC series | parameter | mean | range |
|---|---|---|---|
|  | 1h TEC | 18.3 | 1.8 … 63.8 |
| observed | 1d mean TEC | 18.3 | 8.7 … 33.8 |
|  | 1d max TEC | 30.0 | 13.7 … 63.8 |
|  | 1h TEC | **18.8** | **0.6 … 65.9** |
| PCA-MRM(L31, mean.lag1.2) | 1d mean TEC | 18.8 | 7.6 … **34.2** |
|  | 1d max TEC | **29.4** | **14.1 … 65.9** |
|  | 1h TEC | 17.7 | 0.5 … 68.7 |
| PCA-MRM(L32, mean.lag1.2) | 1d mean TEC | **18.7** | **8.6 … 34.8** |
|  | 1d max TEC | 29.2 | **14.1 … 68.7** |

**Table 2.** Performance the PCA-MRM(L##, mean.lag1.2) and the naïve models for the TEC series. Best scores over the models for each of the studied TEC series (*1h*, *1d mean* and *1d max*) are in bold.

| Metric | PCA-MRM(L31, mean.lag1.2) | | | PCA-MRM(L32, mean.lag1.2) | | | Naïve model |
|---|---|---|---|---|---|---|---|
|  | *1h* | *1d mean* | *1d max* | *1h* | *1d mean* | *1d max* | *1d mean* |
| minimum ΔTEC, TECu | **-14.7** | **-11.9** | -27.8 | **-14.7** | **-11.9** | -25.8 | -11.4 |
| maximum ΔTEC, TECu | 10.7 | **15.3** | 34.0 | **6.7** | 16.6 | **26.7** | 7.9 |
| median ΔTEC, TECu | 0.27 | 0.23 | **-0.35** | 0.22 | 0.18 | -0.45 | 0.64 |
| mean ΔTEC (ME), TECu | **2.1** | 0.22 | **-0.67** | 2.2 | 0.15 | -0.83 | 0.25 |
| r | **0.89** | **0.88** | **0.80** | 0.88 | **0.88** | 0.79 | 0.86 |
| ExpV | **0.78** | **0.77** | **0.63** | 0.77 | **0.77** | 0.62 | 0.74 |
| $R^2$ | **0.80** | **0.77** | **0.63** | 0.79 | **0.77** | 0.61 | 0.73 |
| MAE, TECu | 1.80 | **2.0** | **4.2** | **1.76** | **2.0** | 4.3 | 2.5 |
| MaxE, TECu | **44.7** | 15.3 | 34.0 | 45.4 | 16.6 | **26.7** | 11.4 |
| RMSE, TECu | **4.27** | **2.8** | **5.9** | 4.32 | 2.9 | 6.0 | 3.0 |

Another standard way to assess a new model quality is to compare its forecasts to predictions made by a so-called naïve model, a model in which minimum manipulation of data are used to prepare a forecast. One of the widely used naïve models simply uses the mean value of the studied parameter for a certain time interval as a forecasted value. Here we present a comparison of the PCA-MRM models to the naïve model calculated as averaging of the 1d mean TEC series for the previous 31 days to calculate corresponding forecasts. The last column of Tab. 2 shows metric calculated for the forecasted TEC series using the naïve model. As one can see, the PCA-MRM models outperform the naïve model except for the MaxE, max ΔTEC and min ΔTEC metric. The naïve model constructed in the same way for the 1h TEC series (not shown), in general, forecasts lower TEC values than the observed one for the daytime and higher than the observed TEC values for the night-time, and the metric values are worser than those for the PCA-MRM models.

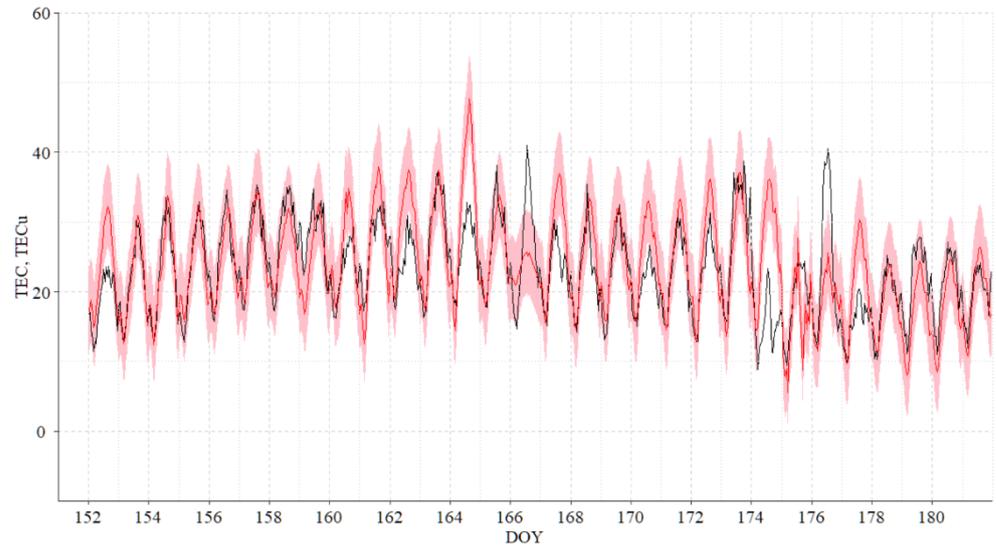

**Figure 4.** Observed (black) and forecasted using the PCA-MRM(L31, mean.lag1.2) model (red) 1h TEC series for June 2015; pink area shows 90% confidence interval (larger version can be found in SM, Fig. S8).

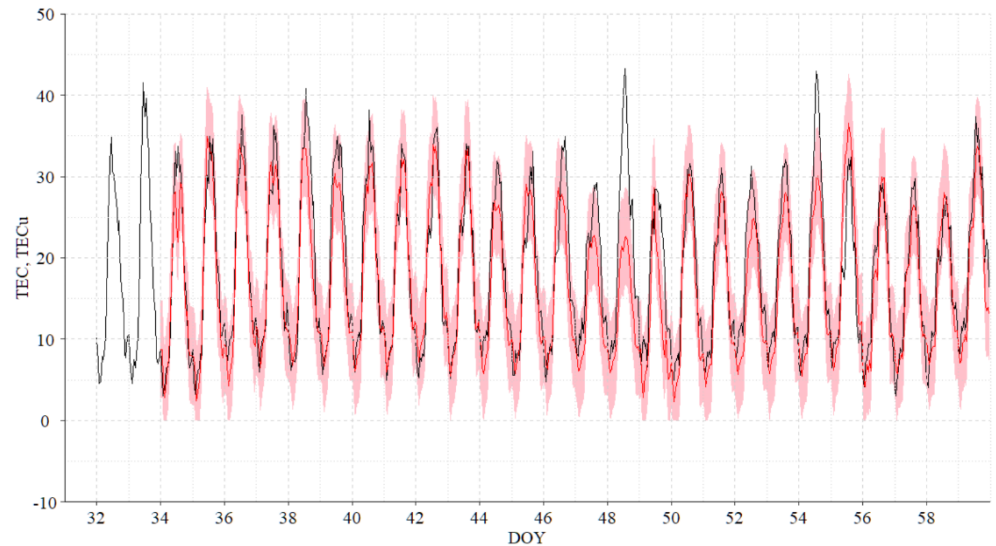

**Figure 5.** Same as Figure 4 but for February 2015 (larger version can be found in SM, Fig. S4).

*5.3. Seasonal variations of the model's performance*

The correlation between the observed and forecasted 1h TEC series is generally high: r = 0.72-0.92, depending on a month with worst correlation obtained for June (Fig. 4 and Fig. S8) and the best correlation obtained for February (r = 0.92, Fig. 5 and Fig. S4) and October of 2015 (r = 0.91, Fig. S12) – see Tab.3. The differences between the observed and forecasted 1h TEC series changes from day to day: in July and December MAE < 10 TECu for all days of a month; in February, March, May, September, October and November there were 1-2 days per month with 10 TECu < MAE < 15 TECu; in April, June and August there were 3-6 days per month with 10 TECu < MAE < 15 TECu; in February and April there was one day per month and in March there were 2 days with MAE ≈ 20 TECu.

To estimate the quality of the PCA-MRM forecasts on the monthly time scale the monthly means were calculated for SWp (both those that were used as predictors and others) and for a number of TEC parameters: 1d mean and 1d max TEC and ΔTEC values (both observed and forecasted by the PCA-MRM models), and the 1h TEC forecasted series (they may be seen in Fig. S15 in SM). The monthly means of the 1d mean and 1d max ΔTEC values show strong anti-correlation with the monthly means of the solar UV (F10.7 and Mg II) and XR fluxes. This anti-correlation results from the underestimation of the amplitude of the TEC daily variations during time intervals with high level of the solar UV and XR irradiance. This underestimation of the TEC daily mean and max values, at least partly, is connected to the underestimation of the flares effect on the daily TEC values, but corresponding correlation coefficients between ΔTEC and the number of the solar flares (Fig. S15 in SM) are low in the absolute values and statistically non-significant.

Finally, the analysis of the correlation coefficients between the observed and forecasted 1h TEC series calculated for individual months (Tab. 3) in dependence on the mean level of the solar, interplanetary and geomagnetic activity level (see also Fig. S15, bottom row) shows that the PCA-MRM models provide best forecasts for months with lowest disturbance level: (1) the months with highest Dst values (Dst ≈ 0 nT), which are the months with lowest geomagnetic activity levels, and/or (2) the months with low number of flares. It is interesting to note that the correlation coefficients obtained for March and October of 2015, months with strong geomagnetic storms [see also 20], are quite high (r = 0.83 and 0.91, respectively), but for June 2015, another month with a strong storm, the correlation is low (r = 0.72). Thus, it seems that what is more important for the PCA-MRM model's performance is not short-living events like storms but the general level of the solar and geomagnetic activity during those *L* days that are used to build a forecast for a particular day.

**Table 3.** Correlation between the PCA-MRM(L31, mean.lag1.2) forecasted and the observed TEC values for different months.

|  | months | | | | | | | | | | |
| --- | --- | --- | --- | --- | --- | --- | --- | --- | --- | --- | --- |
|  | 2 | 3 | 4 | 5 | 6 | 7 | 8 | 9 | 10 | 11 | 12 |
| r | 0.92 | 0.83 | 0.86 | 0.83 | 0.72 | 0.83 | 0.81 | 0.80 | 0.91 | 0.87 | 0.87 |
| p values | <0.01 | <0.01 | <0.01 | <0.01 | <0.01 | <0.01 | <0.01 | <0.01 | <0.01 | <0.01 | <0.01 |

*5.4. General performance of the hourly time scale*

On the hourly time scale the ΔTEC values forecasted by the PCA-MRM(L31, mean.lag1.2) model are in the range from -27.9 TECu to 44.7 TEC for individual days with mean ME = 2.1 ± 0.0005 TECu. The MAE values are in the range from 0 TECu to 44.7 TEC with mean MAE = 1.8 ± 0.0004 TECu.

The daytime hours were defined as the hours from 09:00 UTC to 19:00 UTC and the night-time hours were defined as the hours from 00:00 UTC to 07:00 UTC and from 21:00 UTC to 23:00 UTC of each day independently on the month. The gaps of 2 hours between the day- and night-hours subsets was chosen to avoid the influence of the seasonal variation of the sunrise and sunset times and, consequently, of the phase of the daily TEC variation.

For the whole number of the days with forecast by the PCA-MRM models, MAE = 3.9 ± 0.1 TECu for the day hours (ME =0.5 ± 0.1 TECu), and MAE = 2 ± 0.03 TECu for the night hours (ME = -0.1 ±0.05 TECu) or twice smaller than for the daytime.

**Table 4.** The standard deviations of the observed and forecasted TEC, and the standard deviation of ΔTEC and the mean errors for the PCA-MRM(L31, mean.lag1.2) model: all hours, day hours and night hours.

|  | all hours | day hours | night hours |
| --- | --- | --- | --- |
| SD$_{TEC\ obs}$, TECu | 6.45 ± 0.15 | 3.9 ± 0.10 | 2.50 ± 0.06 |
| SD$_{TEC\ forec}$, TECu | 6.54 ± 0.14 | 3.6 ± 0.10 | 2.20 ± 0.06 |

| | | | | |
|---|---|---|---|---|
| SD$_{\Delta TEC}$, TECu | 2.90 ± 0.09 | 2.9 ± 0.09 | | 1.78 ± 0.05 |
| MAE, TECu | 2.90 ± 0.10 | 3.9 ± 0.16 | | 2.00 ± 0.07 |
| RMSE, TECu | 3.70 ± 0.12 | 4.5 ± 0.17 | | 2.40 ± 0.07 |
| ME, TECu | 0.20 ± 0.15 | 0.5 ± 0.24 | | -0.10 ± 0.11 |
| MaxE, TECu | 8.20 ± 0.25 | 7.9 ± 0.25 | | 4.50 ± 0.14 |

Table 4 shows SD and other metrics of the observed and forecasted, PCA-MRM(L31, mean.lag1.2), TEC series calculated for all forecasted days for all hours and separately for the day and night hours. For the PCA-MRM(L31, mean.lag1.2) model we have forecasts for 332 days of 2015, since the first 33 days of 2015, from January 1 to February 2, are used to build the first possible forecast. As one can see, SDs of the observed and forecasted 1h series calculated using 1h data are very close. MAE is about 3-4 TECu with MaxE ~8 TECu. The errors and SD are lower for the night hours than for the day hours: MAE ~2 TECu vs ~4 TECu and MaxE ~4.5 TECu vs ~8 TECu, respectively.

Table 5 presents the daily $R^2$ and ExpV scores for the PCA-MRM(L31, mean.lag1.2) model: it shows per cent of days with $R^2$ and ExpV above a certain threshold. The scores are calculated for each day for all hours and for the day/night hours separately. Please note that $R^2$ or ExpV can be negative for days with an extremely bad performance of a model when the MSE or $\sigma^2_{\Delta y}$, respectively, are larger than $\sigma_y^2$ (see Eq. 1-2 and 6).

Table 5. Per cent of the forecasted days (out of 332) with $R^2$ and ExpV scores in a certain range (PCA-MRM(L31, mean.lag1.2) model).

| Parameter | Time | threshold | | |
|---|---|---|---|---|
| | | ≥ 0 | ≥ 0.5 | ≥ 0.8 |
| $R^2$ | all hours | 88.0% | 74.7% | 41.9% |
| | day hours | 49.4% | 26.0% | 5.00% |
| | night hours | 57.0% | 29.0% | 7.00% |
| ExpV | all hours | 95.5% | 88.3% | 58.0% |
| | day hours | 76.5% | 48.5% | 28.3% |
| | night hours | 80.0% | 46.4% | 14.8% |

Tables 6-8 show the mean and median values of different error scores (daily mean and median MAE and RMSE, ME, and MaxE) as well as the per cent of days with the scores in a certain range. The last column shows the range of the scores observed for 90% of all forecasted days. The scores are calculated for each day for the whole day and for the day/night hours only. As one can see, in average, the absolute value of the forecast error is ~3 TECu with error of ~2 TECu for the night hours. For all hours, 80-90% of the days have errors with the absolute values not exceeding 5 TECu, whereas for the day hours only ~70% of the days have errors with the absolute values in this range, and almost for all days (95%) the absolute values of the errors for the night hours do not exceed 5 TECu. For 90% of the forecasted days MAE is in the range 0-5 TEC for all hours, 0-7 TECu for the day hours and 0-4 TECu for the night hours. The mean and median values of ME allows to assume that for the day hours the PCA-MRM models tends to overestimate TEC values (ME > 0), whereas for the night hours there is a tendency to the underestimation (ME < 0).

Table 6. Same as Table 5 but for MAE and RMSE.

| Parameter | Time | Mean value, TECu | Median value, TECu | Range, TECu | | | For 90% of the days metric is in the range, TECu |
|---|---|---|---|---|---|---|---|
| | | | | 0...3 | 0...4 | 0...5 | |
| MAE | all hours | 2.9 | 2.6 | 61.7% | 80.0% | 91.5% | 0…4.7 |
| | day hours | 3.9 | 3.2 | 46.0% | 63.0% | 77.7% | 0…6.5 |

|  | night hours | 2 | 1.7 | 80.0% | 93.0% | 95.0% | 0…3.7 |
|---|---|---|---|---|---|---|---|
| RMSE | all hours | 3.7 | 3.3 | 43.7% | 65.7% | 82.2% | 0…5.6 |
|  | day hours | 4.5 | 2.7 | 32.5% | 54.0% | 67.8% | 0…7.3 |
|  | night hours | 2.4 | 2.1 | 73.2% | 86.0% | 95.0% | 0…4.3 |

Table 7. Same as Table 5 but for ME.

| Parameter | Time | Mean value, TECu | Median value, TECu | Range, TECu | | | For 90% of the days metric is in the range, TECu |
|---|---|---|---|---|---|---|---|
|  |  |  |  | -1…1 | -2…2 | -3…3 |  |
| ME | all hours | 0.2 | 0.13 | 36.7% | 59.3% | 77.0% | -5.0…5.0 |
|  | day hours | 0.5 | 0.58 | 31.6% | 48.0% | 62.0% | -6.0…6.0 |
|  | night hours | -0.1 | -0.03 | 58.0% | 77.0% | 85.0% | -3.5…3.5 |

Table 8. Same as Table 5 but for MaxE.

| Parameter | Time | Mean value, TECu | Median value, TECu | Range, TECu | | | For 90% of the days metric is in the range, TECu |
|---|---|---|---|---|---|---|---|
|  |  |  |  | <1 | 1…5 | 1…10 |  |
| MaxE | all hours | 8.2 | 7.2 | 0.30% | 22.0% | 75.0% | 1…13.5 |
|  | day hours | 7.9 | 7 | 0.30% | 26.5% | 77.7% | 1…13.5 |
|  | night hours | 4.5 | 4 | 0.90% | 65.0% | 97.0% | 1…8.00 |

MaxE is lower than 5 TECu for ~25% of the day hours and 65% of the night hours, and for 90% of the days it is lower than 13.5 TECu for the daytime and 8 TECu for the night-time.

For the 1h TEC series 90% of the forecasted values have ΔTEC in the range ± 6 TECu, 90% of the forecasted 1d max values have ΔTEC in the range from -9 to 8 TECu, and 90% of the forecasted 1d mean values have ΔTEC in the range ± 4 TECu. Figures 4-5 and S4-S14 in SM show PCA-MRM(L31, mean.lag1.2) model forecasts with 90% confidence interval (pink area) for the 1h TEC series for individual months from February to December and Figs. 2-3 show similar forecasts for the daily mean and daily max TEC series, respectively. As one can see, the observed TEC series (black lines) fit very well into the 90% confidence intervals of the PCA-MRM model.

*5.5. Assesment of the forecasting skills during quiet days*

The disturbed days (DD) were defined as days with daily mean values of the geomagnetic indices above/below a certain threshold (daily mean Dst ≤ -40 nT, and/or ap ≥ 40, and/or Kp ≥ 4.5) and/or with at least 3 solar flares of the C or M classes. All other days were considered as quiet days (QD).

In general, for QDs the daily ME during the daytime is between 0.6 and 0.95 TECu and during the night-time it is between -0.02 and 0.41 TECu; the MAE values are ~3.4 TECu for the daytime and are ~2 TECu during the night-time. Thus, during QDs the amplitude of MAE during the night hours is ~1.7 times smaller than during the day hours.

To study the effect of the slightly elevated solar or geomagnetic activity, we made following QD subsets (see Tab.9): the subsets from QD1 to QD3 show the effect of 1-2 solar flares on the day-night differences of ΔTEC, and the subsets QD1 and from QD4 to QD7 show the effect of the elevated geomagnetic activity. As one can see from Tab. 9, the increase of the daily number of the C flares from 0 to 2 results in the decrease of the mean ΔTEC values for the night hours but the mean MAE values remain constant. This means that the PCA-MRM models tend to underestimate TEC values for the night hours of QD with 1 or 2 flares of up to C-class. For the day hours no statistically significant difference in ΔTEC or MAE is obtained. An increase of the geomagnetic activity during QD seems to result in the overestimation by the PCA-MRM models of the night TEC values: ΔTEC increases, MAE does not change.

**Table 9.** Parameters of the subsets of QDs and the mean ΔTEC (in TECu) values. Data are for the PCA-MRM model with *L* = 31 days and averaging the model with lag = 1 day and lag = 2 days.

| Subset number | Kp | ap | Dst, nT | Number of C flares | Number of days in the subset | day | | night | |
|---|---|---|---|---|---|---|---|---|---|
| | | | | | | ΔTEC | \|ΔTEC\| | ΔTEC | \|ΔTEC\| |
| QD1 | < 30 | < 20 | > -40 | 0 | 51 | 0.95 ± 0.17 | 3.22 ± 0.11 | 0.32 ± 0.11 | 1.95 ± 0.07 |
| QD2 | < 30 | < 20 | > -40 | 0-1 | 113 | 0.64 ± 0.12 | 3.3 ± 0.08 | 0.07 ± 0.07 | 1.93 ± 0.05 |
| QD3 | < 30 | < 20 | > -40 | 0-2 | 144 | 0.8 ± 0.11 | 3.3 ± 0.07 | -0.02 ± 0.06 | 1.9 ± 0.04 |
| QD4 | < 20 | < 10 | > -40 | 0 | 37 | 0.93 ± 0.2 | 3.3 ± 0.12 | 0.25 ± 0.13 | 2.06 ± 0.09 |
| QD5 | < 40 | < 30 | > -50 | 0 | 59 | 0.81 ± 0.18 | 3.5 ± 0.13 | 0.33 ± 0.1 | 1.96 ± 0.07 |
| QD6 | < 50 | < 40 | > -60 | 0 | 61 | 0.87 ± 0.18 | 3.5 ± 0.12 | 0.37 ± 0.1 | 2 ± 0.07 |
| QD7 | < 60 | < 50 | > -70 | 0 | 64 | 0.91 ± 0.17 | 3.5 ± 0.12 | 0.41 ± 0.1 | 1.99 ± 0.07 |

*5.6. Assesment of the forecasting skills during space weather events*

In general, for DDs the daily ME values during the daytime are between -7.6 and 0.61 TECu and during the night-time they are between -0.73 and 0.54 TECu, the daily MAE values during the daytime are between 3.3 and 12.8 TECu and during the night-time they are ~2.2 TECu. Thus, during DDs the amplitude of MAE during the night hours is ~1.6-4.2 times smaller than during the day hours.

**Table 10.** Parameters of the subsets of DDs and the mean ΔTEC (in TECu) values. Data are for the PCA-MRM model with *L* = 31 days and averaging the model with lag = 1 day and lag = 2 days.

| Subset number | Kp | ap | Dst | Number of flares | Days in the subset | day | | night | |
|---|---|---|---|---|---|---|---|---|---|
| | | | | | | ΔTEC | \|ΔTEC\| | ΔTEC | \|ΔTEC\| |
| | | | no geomagnetic disturbance but solar flares | | | | | | |
| DD1 | < 30 | < 20 | > -40 | > 2 | 86 | -0.04 ± 0.16 | 3.8 ± 0.1 | -0.53 ± 0.09 | 2.07 ± 0.06 |
| DD2 | < 30 | < 20 | > -40 | > 3 | 62 | 0.09 ± 0.19 | 3.6 ± 0.12 | -0.5 ± 0.11 | 2 ± 0.07 |
| DD3 | < 30 | < 20 | > -40 | > 4 | 38 | -0.44 ± 0.21 | 3.3 ± 0.14 | -0.69 ± 0.13 | 2 ± 0.09 |
| | | | geomagnetic disturbance without solar flares | | | | | | |
| DD4 | > 30 | > 20 | < -40 | < 1 | 8 | -0.48 ± 0.92 | 5.8 ± 0.69 | 0.54 ±0.31 | 2.1 ± 0.22 |
| DD5 | > 40 | > 30 | < -50 | < 1 | 4 | -3.3 ± 1.6 | 7.7 ± 1.2 | -0.24 ± 0.47 | 2 ± 0.36 |
| DD6 | > 50 | > 40 | < -60 | < 1 | 2 | -7.6 ± 3 | 12.8 ± 1.9 | -0.73 ± 0.9 | 3 ± 0.63 |
| | | | geomagnetic disturbance and/or solar flares | | | | | | |
| DD7 | > 40 | > 30 | < -50 | > 2 | 161 | 0.36 ± 0.15 | 4.4 ± 0.11 | -0.14 ± 0.07 | 2.1 ± 0.05 |
| DD8 | > 50 | > 40 | < -60 | > 2 | 155 | 0.37 ± 0.15 | 4.4 ± 0.11 | -0.13 ± 0.07 | 2.1 ± 0.05 |
| DD9 | > 50 | > 40 | < -60 | > 3 | 121 | 0.53 ± 0.18 | 4.5 ± 0.13 | -0.11 ± 0.08 | 2.1 ± 0.06 |
| DD10 | > 50 | > 60 | < -100 | > 3 | 112 | 0.61 ± 0.19 | 4.43 ± 0.14 | -0.16 ± 0.08 | 2.1 ± 0.06 |

To study the effect of the solar flares and geomagnetic disturbances separately, we made following DD subsets (Tab. 10): the subsets DD1 to DD3 show the effect of the solar flares on the day-night differences of ΔTEC, and the subsets DD4 to DD6 show the effect of the geomagnetic activity (please note the low number of the days with geomagnetic disturbances but without the solar flares). The subset DD7-DD10 contains days either with at least one of the geomagnetic indices above (below for Dst) a threshold and with at least 3 solar flares of any type.

As one can see from Tab. 10, the increase of the daily number of the flares above 2 per day on a geomagnetically quiet background (DD1-DD3) does not result in a systematic under- or over-estimation by the PCA-MRM models of the daytime ΔTEC ≈ 0 TECu (MAE

≈ 3 TECu), but the night-time ΔTEC are underestimated by ~0.5 TECu (MAE ≈ 2 TECu). In turn, the geomagnetic disturbances that are not accompanied by solar flares (DD4-DD6) seems to result in the underestimation by the PCA-MRM models of both the day and night ΔTEC values (ΔTEC tends to be negative). The amplitude of the differences between the forecast and the observations (|ΔTEC|) increases with the strength of a disturbance mostly for the day hours. Still these conclusions are made on the low number of the days. Finally, when either geomagnetic disturbance (seen at least in one of the indices) or more than 2 solar flares are observed during a studied day (DD7-DD10), the daytime ΔTEC is overestimated and the night-time ΔTEC is underestimated by the PCA-MRM models with the MAE ≈ 4 TECu for the day hours and ~2 TECu for the night hours.

For all days of 2015 there is either very weak or no correlation between ΔTEC and the geomagnetic indices: for Dst $r = -0.2$ (p value < 0.01), for ap and Kp $r = 0$. On the other hand, if we compared the geomagnetic indices to MAE, there is relatively weak but statistically significant dependence of the |ΔTEC| values on the strength of geomagnetic disturbances: for Dst $r = -0.35$ (p value < 0.01), for ap $r = 0.32$ (p value < 0.01) and for Kp $r = 0.25$ (p value < 0.01). Thus, for the geomagnetically disturbed days the forecasting quality of the PCA-MRM models decreases, MAE increases, but the under- and over-estimations of the daily mean TEC are observed with more or less similar frequencies.

For the geomagnetically disturbed days, the dependence of ΔTEC and |ΔTEC| on the geomagnetic activity are different (see Tab. 11), however not all correlations are statistically significant due to the low number of corresponding events. For Dst the correlation coefficients obtained for |ΔTEC| is higher in the absolute values than the correlation coefficients obtained for ΔTEC showing, again, that the over- and under-estimations of the daily mean TEC are observed with similar frequencies.

**Table 11.** Correlation coefficients r between ΔTEC and SW parameters observed during non-quiet days: days with larger number of flares or geomagnetic disturbances; p values are in parentheses. Statistically significant (95%) **r** are in bold.

| Days with… | Correlation with… | r for ΔTEC | r for |ΔTEC| |
|---|---|---|---|
| Daily mean Dst -40 nT | Dst | **-0.42 (0.03)** | **-0.58 (<0.01)** |
| Daily mean ap ≥ 40 | ap | **-0.46 (0.05)** | 0.36 (0.25) |
| Daily mean Kp ≥ 4.5 | Kp | -0.44 (0.13) | 0.21 (0.55) |

The dependence of ΔTEC and |ΔTEC| on the number of the solar flares was studied separately for flares of different types (C and M flares and the total number of flares per day N). Please note that most of the flares observed in 2015 were of the C-type (951 out of 1018 flares observed in 2015). We found no correlation between ΔTEC or MAE and the number of flares N when all events are considered, however, if we consider only days with at least 5 flares of any type, there is a weak correlation between the number of flares per day and the difference between the observed and forecasted TEC shown in Tab. 12. The correlation coefficients are statistically significant (> 95%) and show that the amplitude of the forecast error (ΔTEC and |ΔTEC|) grows with the number of flares observed per day.

**Table 12.** Correlation coefficients r between ΔTEC and SW parameters observed during non-quiet days: days with larger number of flares or geomagnetic disturbances; p values are in parentheses. Statistically significant (95%) **r** are in bold.

| Days with… | Correlation with… | r for ΔTEC | r for |ΔTEC| |
|---|---|---|---|
| 5 or more flares of all type | Number of flares N | **0.26 (<0.01)** | **0.34 (<0.01)** |
| 5 or more of C flares | Number of C flares | **0.32 (<0.01)** | **0.36 (<0.01)** |
| 2 or more of M flares | Number of M flares | -0.37 (0.18) | -0.26 (0.28) |

The analysis of the ΔTEC and MAE distribution shows that for the days with moderate number of flares (<5) there is a tendency for the PCA-MRM models to underestimate the daily mean TEC values, whereas for the days with larger number of flares (5 or more) there is a tendency for the models to overestimate the daily mean TEC values.

The number of the M flares observed in 2015 (66 flares) is much lower than the number of the C flares. In most cases there was only one flares of this class per day, and only for 13 days of 2015 two or more M-class flares were observed. Therefore, the results of the correlation analysis for these days (Tab. 12, bottom row) are statistically non-significant, and we cannot make a conclusion on the existence of the dependence of ΔTEC on the number of the flares of this type.

Since during 2015 there was only one short-living flare of the X type, no conclusion of the effect of this type flares on the forecasting quality of the PCA-MRM model can be made.

**6. Conclusions**

We propose a PCA-MRM model based on the principal component analysis (PCA) and the multiple linear regression (MRM) of the amplitudes of the PCA modes to forecast TEC. Several space weather parameters are used as regressors for MRMs. The analysis of the performance of this model on the data obtained in 2015 showed that such model can be successfully used to forecasts TEC variations at middle latitudes.

The best length of the input data set is 31 or 32 days. We found that the best performance is seen when the PCA-MRM forecasts made with time lags of 1 and 2 days (space weather parameters lead) are averaged.

The PCA-MRM model allows 90% confidence intervals of 4 TECu for day hours and 2 TECu for night hours, in average. For the quiet days (days without M or X flares and no more than 1 flares of the class C or below, and without geomagnetic disturbances) the MAE and RMSE are about 3-3.5 TECu; for geomagnetically disturbed days without flares MAE and RMSE are about 5-7 TECu; for the days with M or X flares and/or with more than 1 flares of the class C or below MAE and RMSE are about 3.5-4 TECu. Thus, the PCA-MRM model perform well during days without significant geomagnetic disturbances even if a flare is observed.

The daily mean and monthly mean ME and MAE correlate with the solar UV and XR flux: the PCA-MRM model systematically slightly underestimates TEC values for days with high level of the solar UV and XR irradiance. The daily mean and monthly mean ME and MAE correlates with geomagnetic indices Dst, Kp, ap: the PCA-MRM model both under- and over- estimates TEC values during days with geomagnetic disturbances with approximately similar rates however large overestimations are seen more often than large underestimations.

Comparing to other TEC models, both global and regional, the PCA-MRM model based on a single-station TEC data performs very well having RMSE (and other metrics) in the same range as for other models (see Section 1 and [1-12]). We also see a drop in the model's performance during geomagnetically active periods while solar flares alone have no strong effect on the model's performance. Similar to [4 and 12], we found that a time lags of 1 and 2 days between the TEC and SWp series must be introduced for better forecast quality. Also, our results allow to provide different confidence intervals for the day and night hours: the forecast errors for the nigh hours are ~1.6-2 times smaller than ones for the day hours (depending on the level of geomagnetic activity). This is in an agreement with [12].

On the other hand, our model has at least two advantages comparing to many other models. First, we do not need to make assumption on the character of the daily and seasonal TEC variations since the amplitude and the phase of the daily TEC variation for certain time interval (e.g., a month) is extracted automatically from the data used to build a model. Second, our model uses a limited set of the TEC and SWp observations: just 31 or 32 days of data (1h data for TEC and 1d data for SWp) are needed to build the model. Thus, the model is not demanding on the database length (computer memory) and computational time, and, therefore, can be used by small private enterprises (like those participated in the ESA Small ARTES Apps project SWAIR) that monitor ionosphere conditions for GNSS-service consumers as a simple model for a short-term (1 day ahead) regional TEC forecasting.

**Supplementary Materials:** The following are available online at www.mdpi.com/xxx/s1 in a single pdf file: Figure S1: Examples of PC1s and PC2s of the TEC variations; Figures S2-S3: Scores of PCA-MRM models with different *L* values for 1h TEC, 1d meand and 1d max TEC series, respectively; Figures S4-S14: Observed and forecasted using the PCA-MRM(L31, mean.lag1.2) model 1h TEC series for June 2015 with 90% confidence interval for March to May and from July to December of 2015; Figure S15: Correlation between TEC and space weather parameters on the monthly timescale.

**Author Contributions:** Conceptualization, A.M.; Te.B.; Ta.B.; methodology, A.M.; Ta.B.; software, A.M.; Ta.B.; validation, A.M.; Te.B.; Ta.B..; formal analysis, A.M.; Ta.B.; investigation, A.M.; resources, Te.B.; data curation, A.M.; Ta.B.; writing—original draft preparation, A.M.; writing—review and editing, Te..B.; Ta.B.; visualization, A.M.; project administration, Te.B.; funding acquisition, Te.B. All authors have read and agreed to the published version of the manuscript." Please turn to the CRediT taxonomy for the term explanation. Authorship must be limited to those who have contributed substantially to the work reported.

**Funding:** This research was supported through the project "SWAIR—Space weather impact on GNSS service for Air Navigation," ESA Small ARTES Apps, https://about.swair.ptech.io/. IA is supported by Fundação para a Ciência e a Tecnologia (FCT) through the research grants UIDB/04434/2020 and UIDP/04434/2020.

**Data Availability Statement:**

The TEC data for 2015 are available at Barlyaeva, T.; Barata, T.; Morozova, A.; 2020. Datasets of ionospheric parameters provided by SCINDA GNSS receiver from Lisbon airport area, Mendeley Data, v1 http://dx.doi.org/10.17632/kkytn5d8yc.1

"SCINDA-Iono" toolbox for MATLAB by T. Barlyaeva is available online at https://www.mathworks.com/matlabcentral/fileexchange/71784-scinda-iono_toolbox.

We acknowledge the mission scientists and principal investigators who provided the data used in this research:

The TEC data sets are from the Royal Observatory of Belgium (ROB) data base and are publicly available in IONEX format at ftp://gnss.oma.be/gnss/products/IONEX/, see also [21] for more information.

The Dst index is from the Kyoto World Data Center http://wdc.kugi.kyoto-u.ac.jp/dst_final/index.html.

Geomagnetic data measured by the Coimbra Geomagnetic Observatory are available at the World Data Centre for Geomagnetism web portal http://www.wdc.bgs.ac.uk/dataportal/ .

The solar wind data and the ap index are from the SPDF OMNIWeb database. The OMNI data were obtained from the GSFC/SPDF OMNIWeb interface at https://omniweb.gsfc.nasa.gov, see also [27] for more details.

The Mg II data are from Institute of Environmental Physics, University of Bremen http://www.iup.uni-bremen.de/gome/gomemgii.html, see also [22] for more information.

The data on the variations of the solar XR flux are from the LASP Interactive Solar Irradiance Data Center (LISRD, http://lasp.colorado.edu/lisird/ ). LISIRD provides a uniform access interface to a comprehensive set of Solar Spectral Irradiance (SSI) measurements and models from the soft X-ray (XUV) up to the near infrared (NIR), as well as Total Solar Irradiance (TSI). The XRTIMED data are from the Solar EUV Experiment (SEE) measures the solar ultraviolet full-disk irradiance for the NASA TIMED mission. Level 3 data represent daily averages and are filtered to remove flares available at http://lasp.colorado.edu/lisird/data/timed_see_ssi_l3/.

The X-ray Flare dataset was prepared by and made available through the NOAA National Geophysical Data Center (NGDC). The data about the solar flares for 2015 are from https://www.ngdc.noaa.gov/stp/space-weather/solar-data/solar-features/solar-flares/x-rays/goes/xrs/goes-xrs-report_2015_modifiedreplacedmissingrows.txt.

**Acknowledgments:** We are grateful to SEGAL (Space and Earth Geodetic Analysis laboratory) and personally to Dr. Rui Fernandes from University of Beira Interior (Portugal) for the access to the SCINDA receiver data.

**Conflicts of Interest:** The authors declare no conflict of interest.

# Supplementary Materials

for

# PCA-MRM model to forecast TEC at middle latitudes

by

**Anna L. Morozova [1,*], Teresa Barata [2] and Tatiana Barlyaeva [3]**

1
1 Instituto de Astrofísica e Ciências do Espaço, Univ Coimbra, OGAUC, Coimbra, Portugal; annamorozovauc@gmail.com
[2] Instituto de Astrofísica e Ciências do Espaço, Univ Coimbra, OGAUC, Coimbra, Portugal; mtbarata@gmail.com
[3] Instituto de Astrofísica e Ciências do Espaço, Univ Coimbra, OGAUC, Coimbra, Portugal; tvbarlyaeva@gmail.com
* Correspondence: annamorozovauc@gmail.com

This file contains Supplementary Materials (SM) referenced in the main text:

**Figure S1**: Examples of PC1s and PC2s of the TEC variations (*page 2*).

**Figures S2-S3**: Scores of PCA-MRM models with different L values for 1h TEC, 1d mean and 1d max TEC series, respectively (*pages 3-4*).

**Figures S4-S14**: Observed and forecasted using the PCA-MRM(L31, mean.lag1.2) model TEC 1h series for June 2015 with 90% confidence interval for February to December of 2015 (*pages 5-15*).

**Figure S15**: Correlation between TEC and space weather parameters on the monthly timescale (*page 16*).

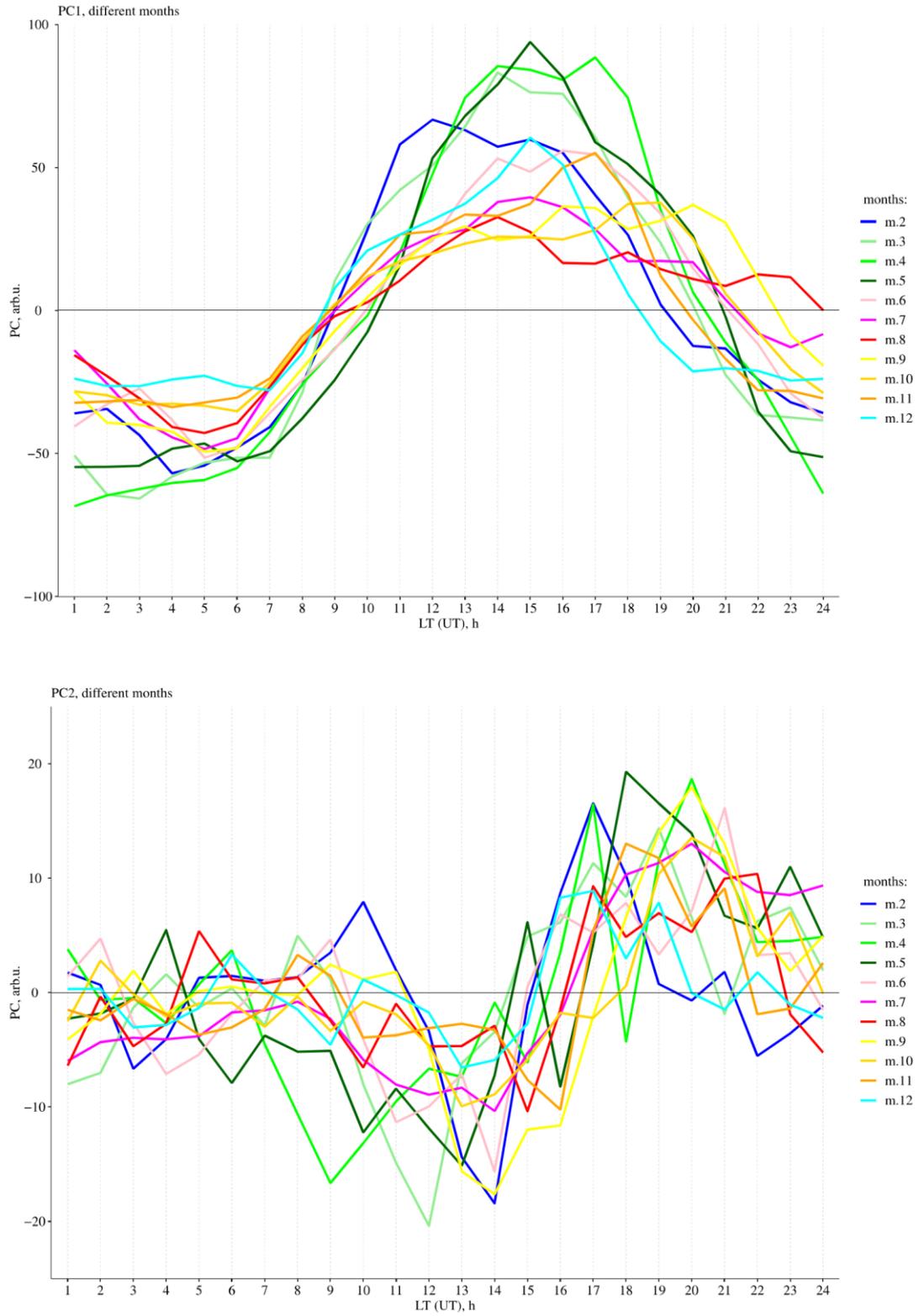

**Figure S1.** Examples of PC1 (top) and PC2 (bottom) of the TEC variations for months from February to December.

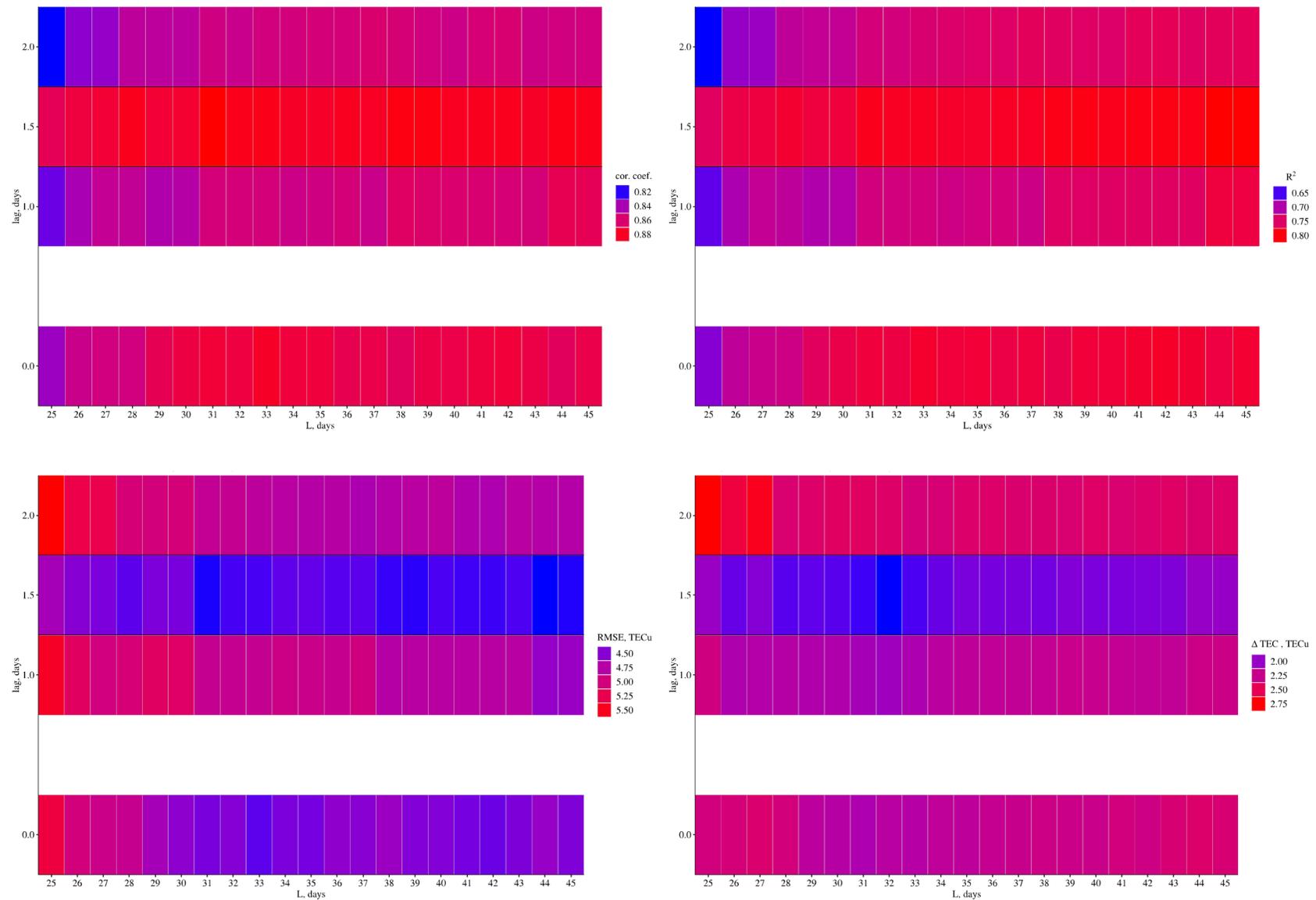

**Figure S2**. Scores of PCA-MRM models with different L and lag values for 1h TEC: correlation coefficients between the observed and forecasted series (top left), $R^2$ (top right), RMSE (bottom left) and MAE (bottom right).

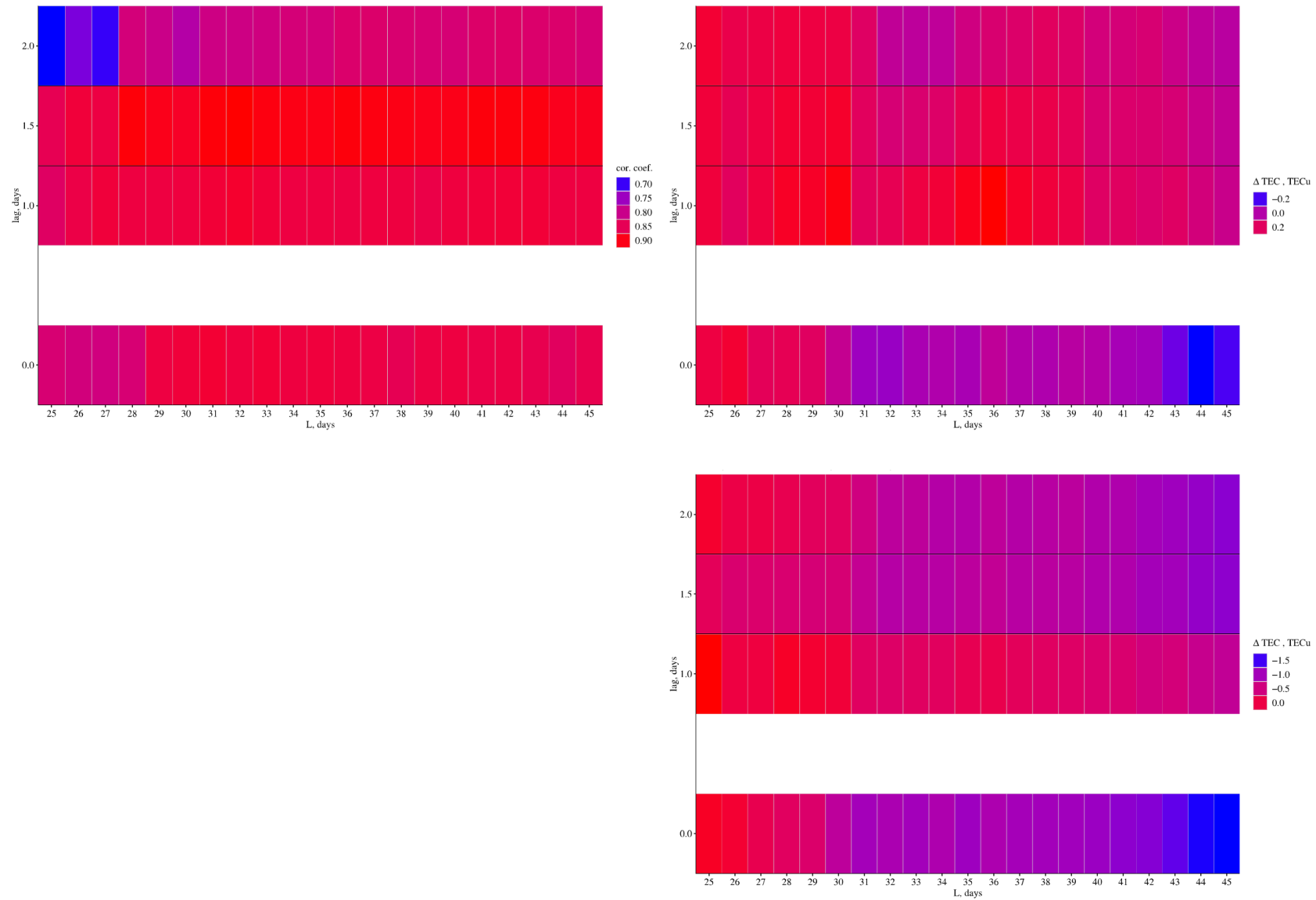

**Figure S3.** Scores of PCA-MRM models with different L and lag values for 1d mean (top) and 1d max TEC (bottom): correlation coefficients between the observed and forecasted series (top left) and MAE (top right and bottom right).

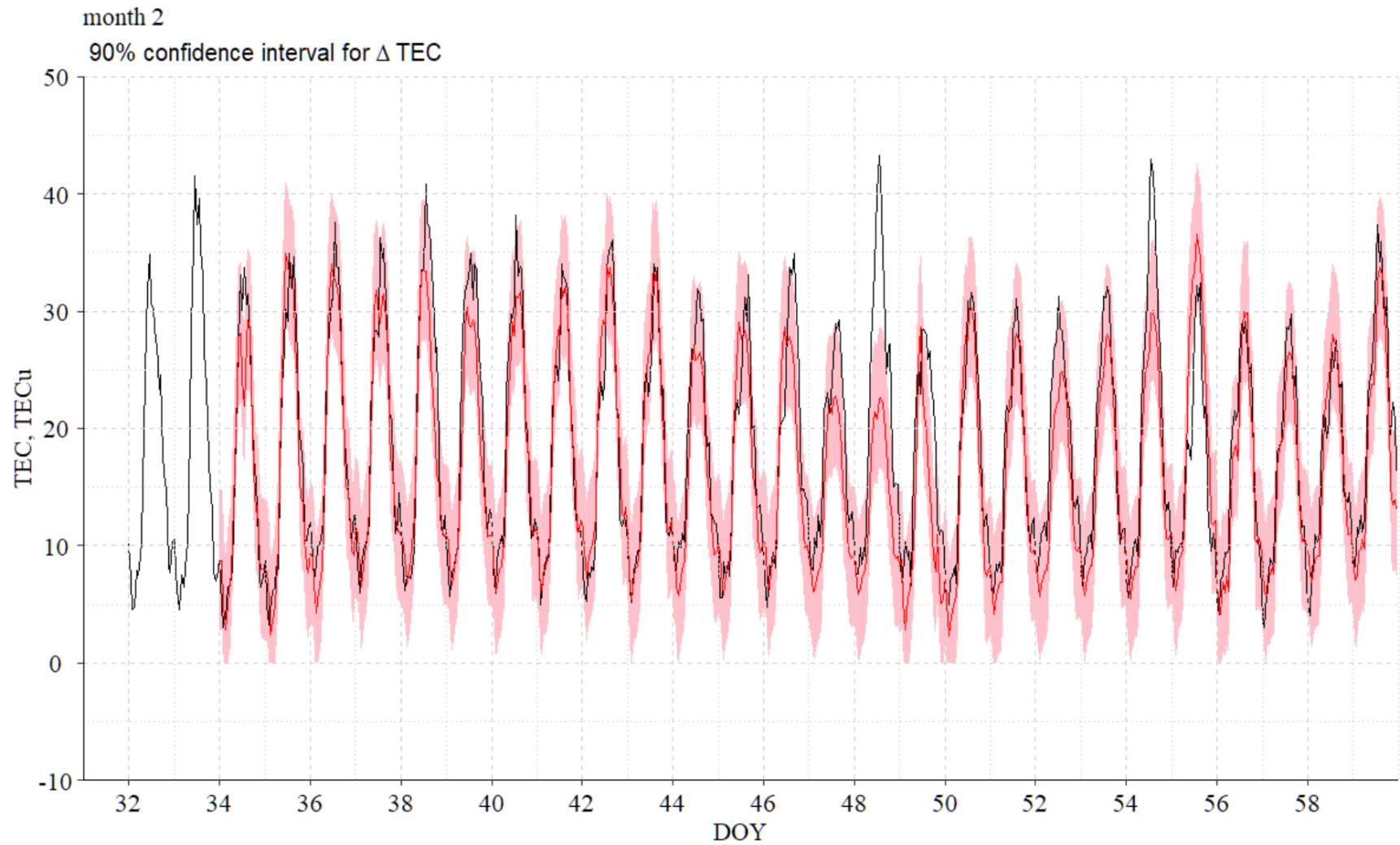

**Figure S4.** Observed (black) and forecasted using the PCA-MRM(L31, mean.lag1.2) model (red) TEC 1h series for March 2015; pink area shows 90% confidence interval.

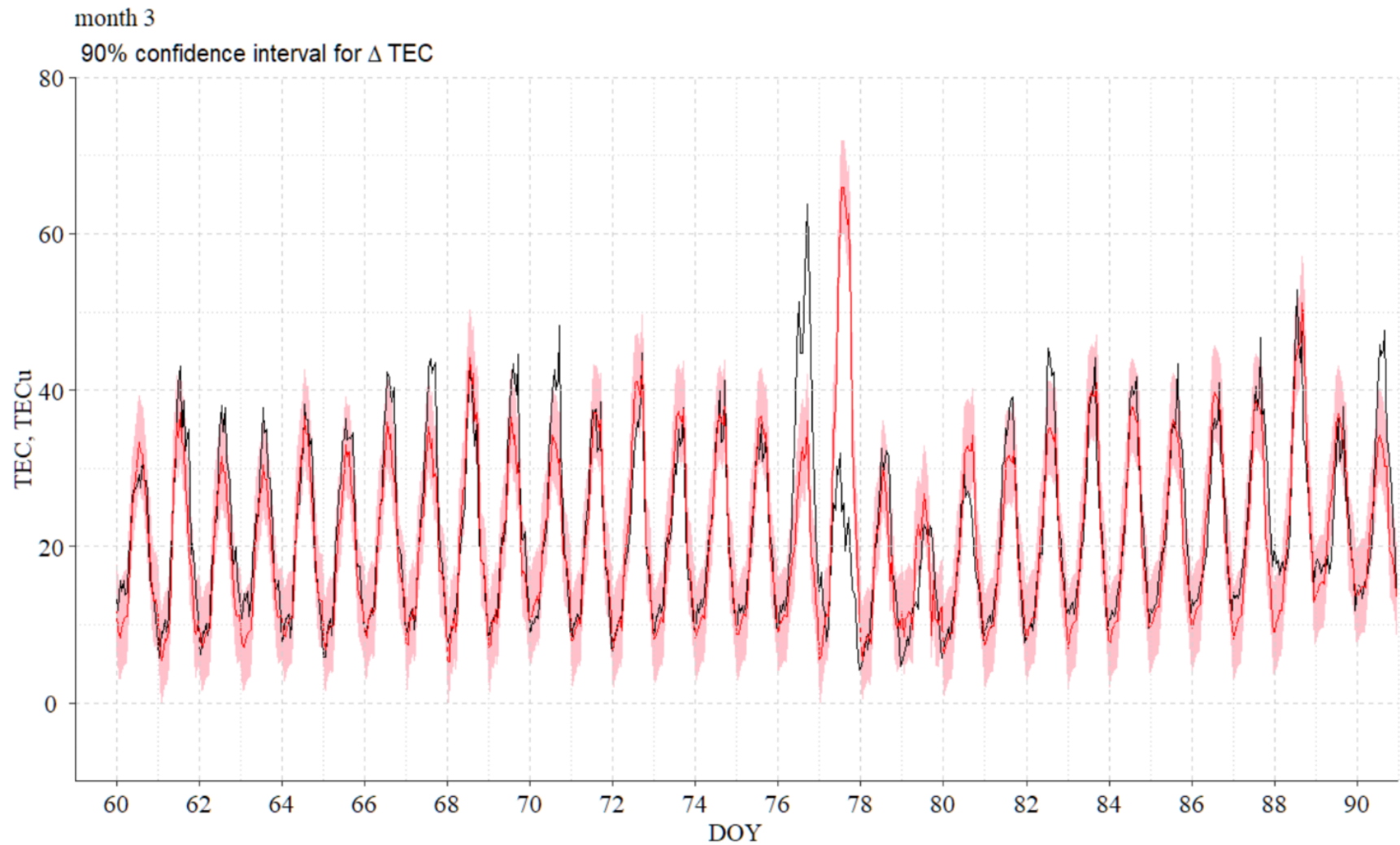

**Figure S5.** Observed (black) and forecasted using the PCA-MRM(L31, mean.lag1.2) model (red) TEC 1h series for March 2015; pink area shows 90% confidence interval.

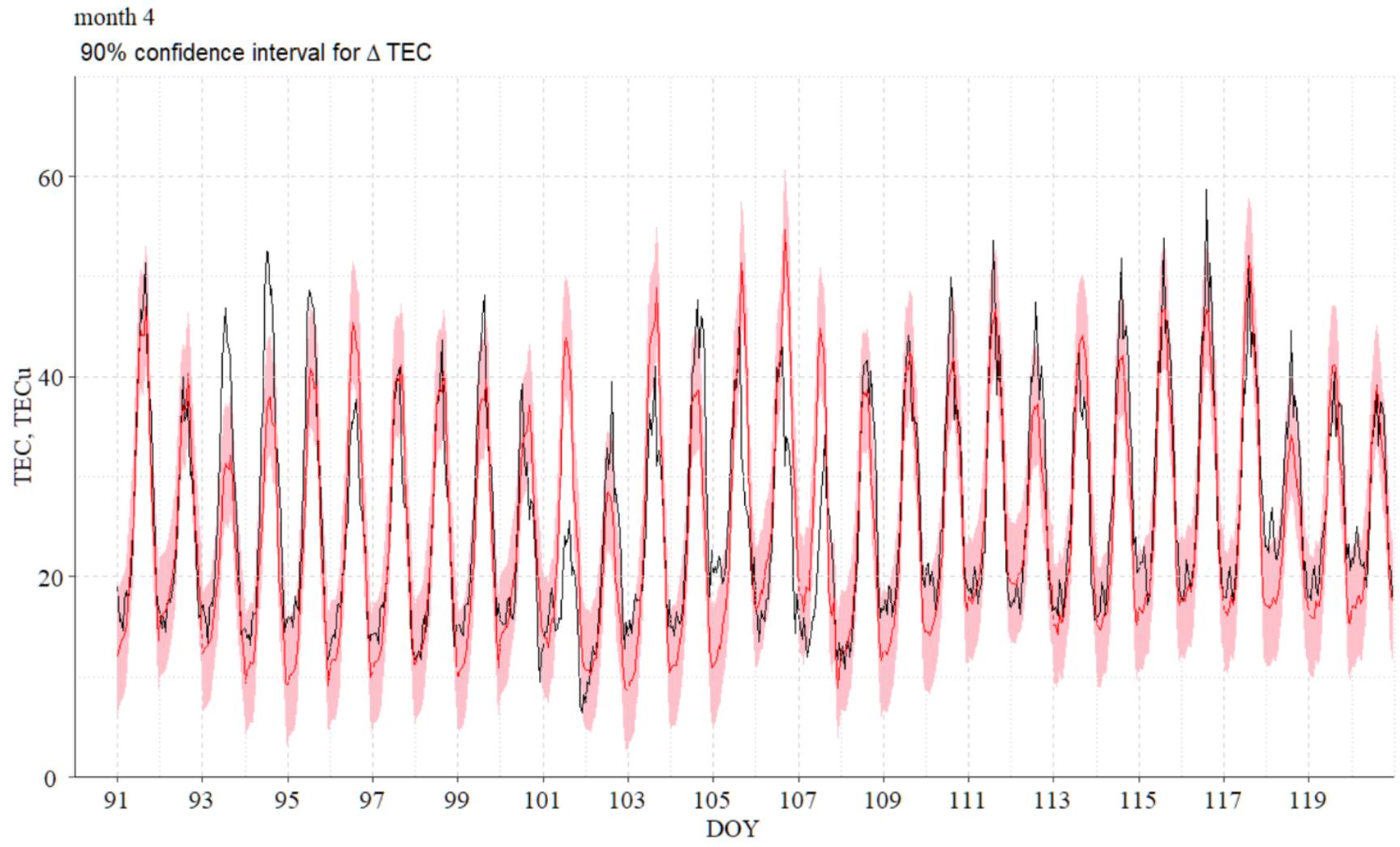

**Figure S6.** Observed (black) and forecasted using the PCA-MRM(L31, mean.lag1.2) model (red) TEC 1h series for April 2015; pink area shows 90% confidence interval.

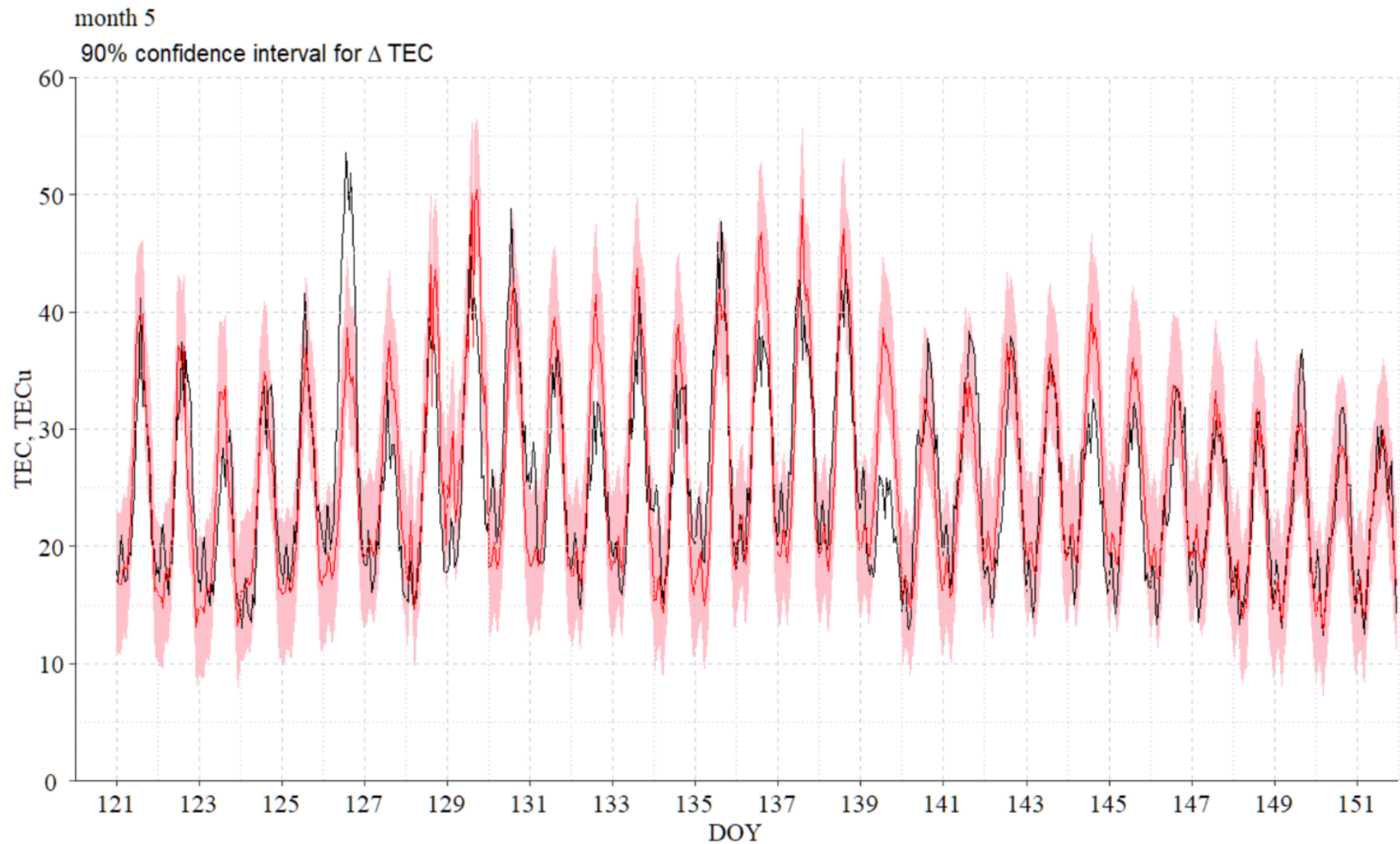

**Figure S7.** Observed (black) and forecasted using the PCA-MRM(L31, mean.lag1.2) model (red) TEC 1h series for May 2015; pink area shows 90% confidence interval.

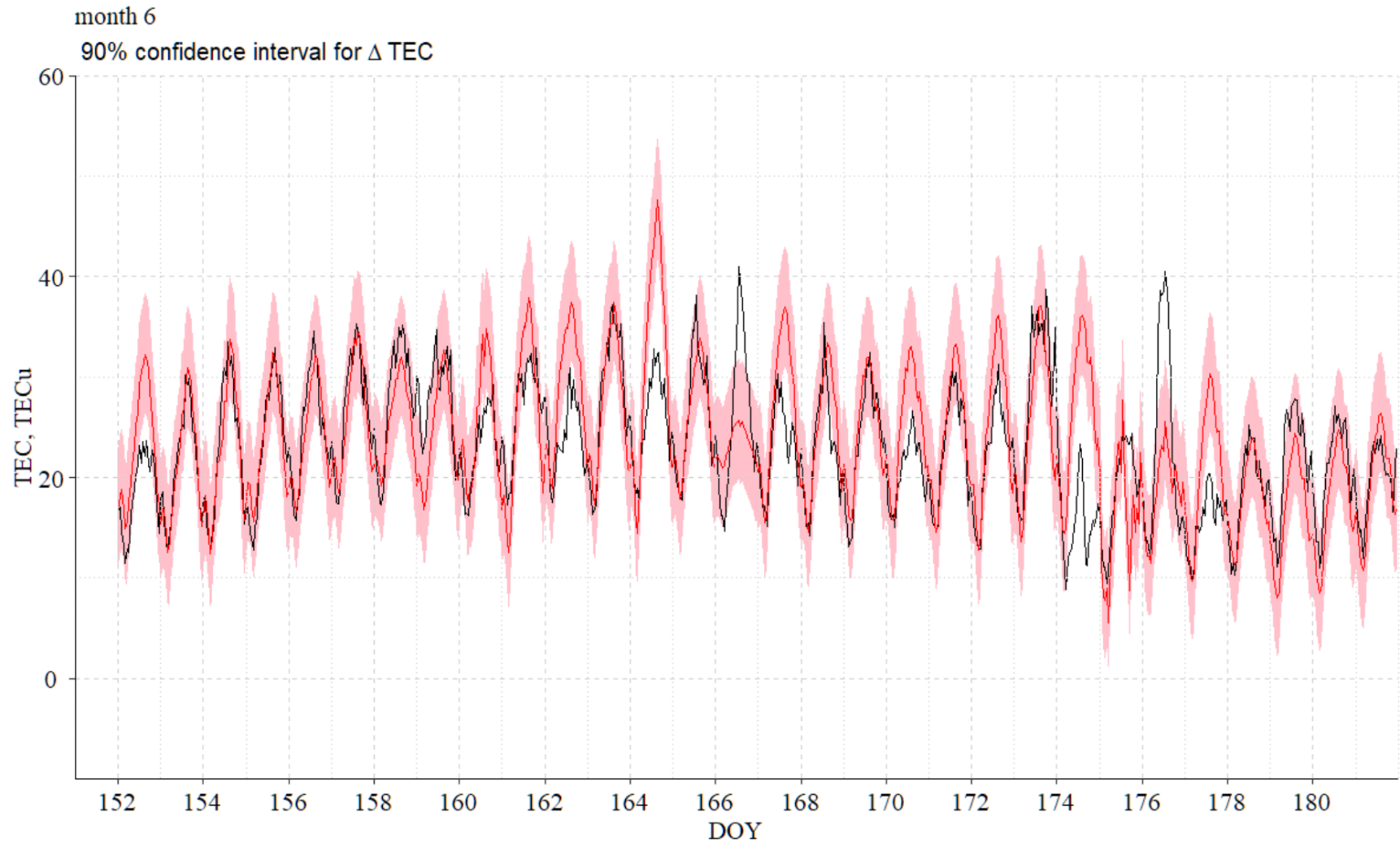

**Figure S8.** Observed (black) and forecasted using the PCA-MRM(L31, mean.lag1.2) model (red) TEC 1h series for March 2015; pink area shows 90% confidence interval.

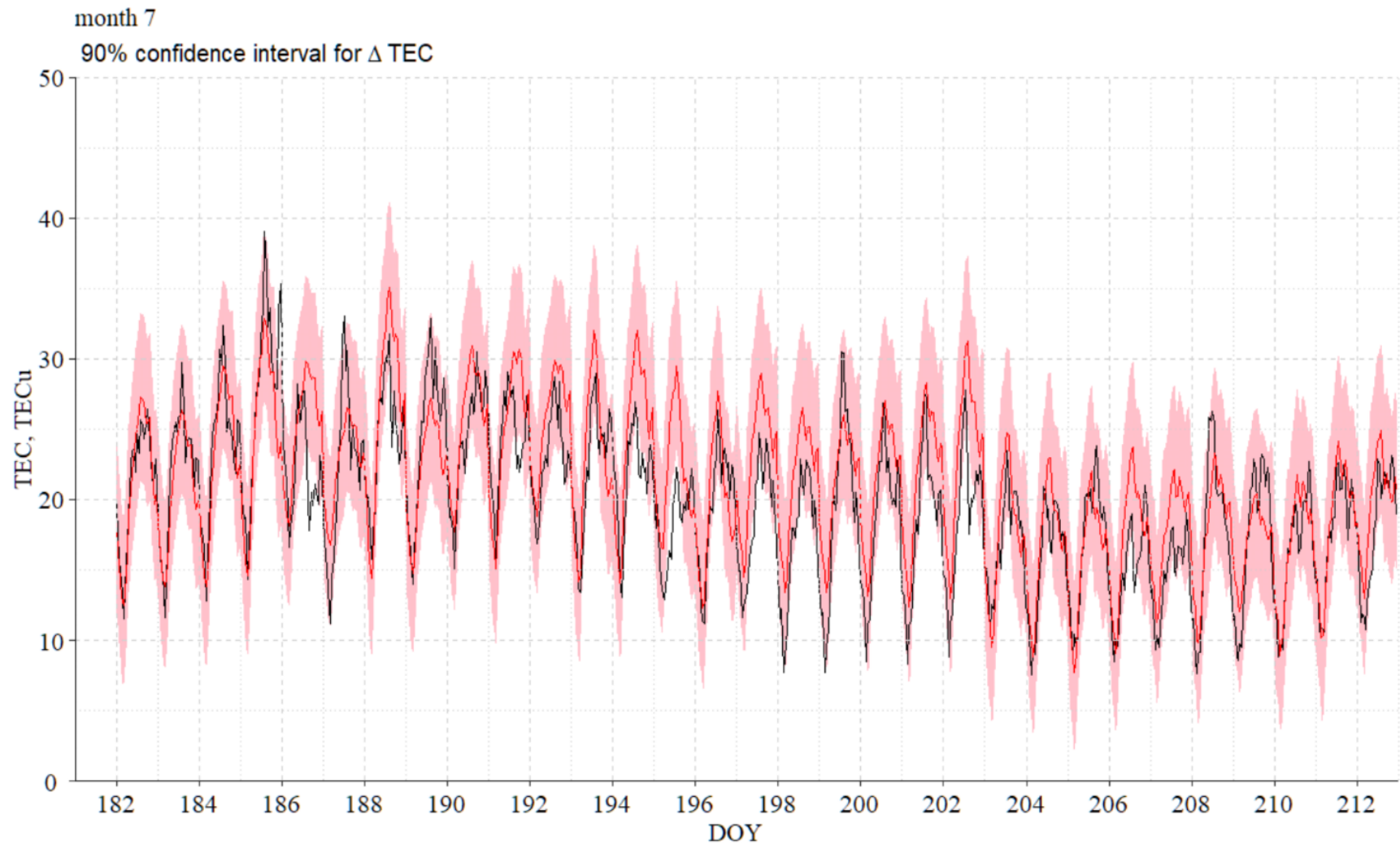

**Figure S9.** Observed (black) and forecasted using the PCA-MRM(L31, mean.lag1.2) model (red) TEC 1h series for July 2015; pink area shows 90% confidence interval.

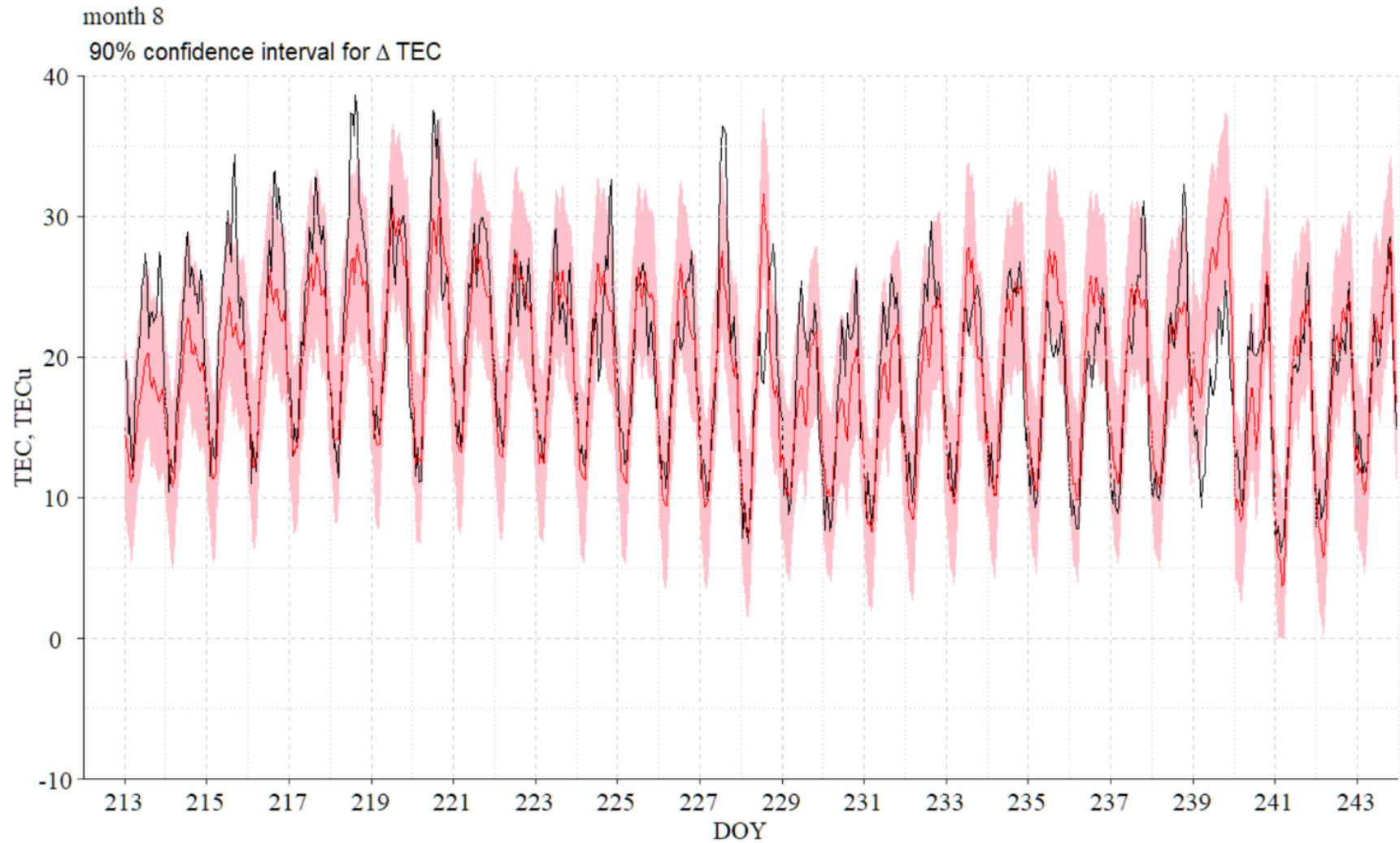

**Figure S10.** Observed (black) and forecasted using the PCA-MRM(L31, mean.lag1.2) model (red) TEC 1h series for August 2015; pink area shows 90% confidence interval.

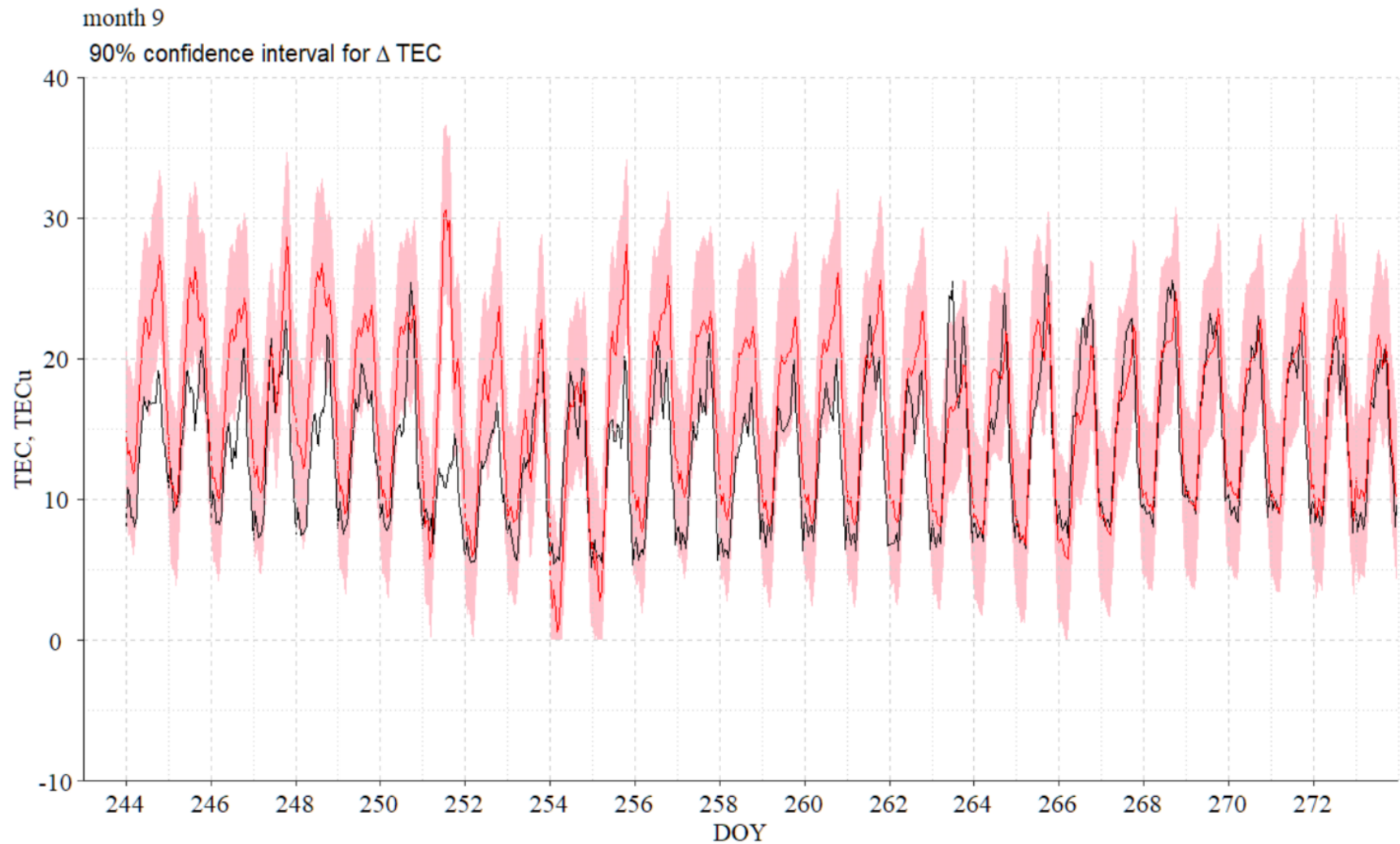

**Figure S11.** Observed (black) and forecasted using the PCA-MRM(L31, mean.lag1.2) model (red) TEC 1h series for September 2015; pink area shows 90% confidence interval.

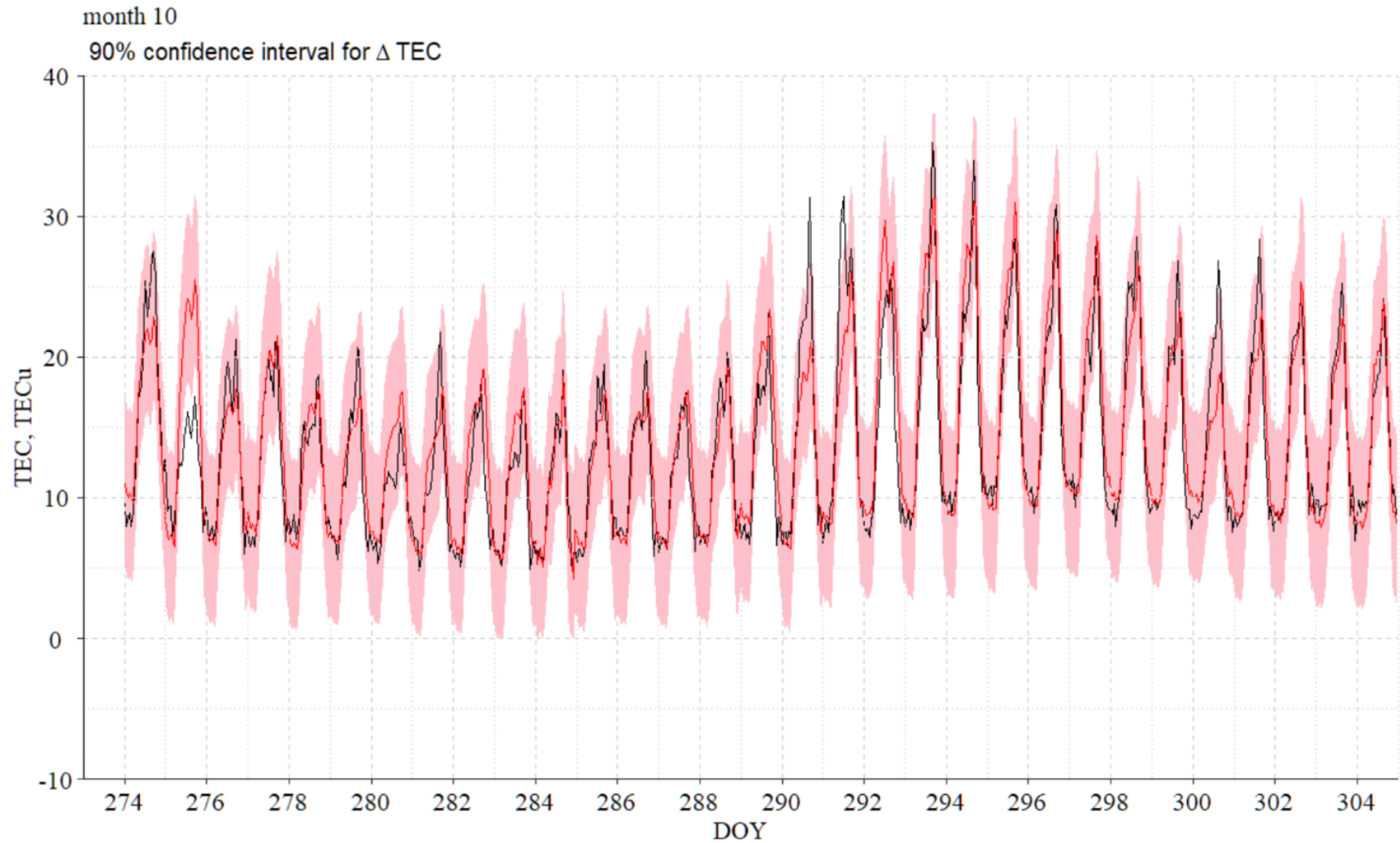

**Figure S12.** Observed (black) and forecasted using the PCA-MRM(L31, mean.lag1.2) model (red) TEC 1h series for October 2015; pink area shows 90% confidence interval.

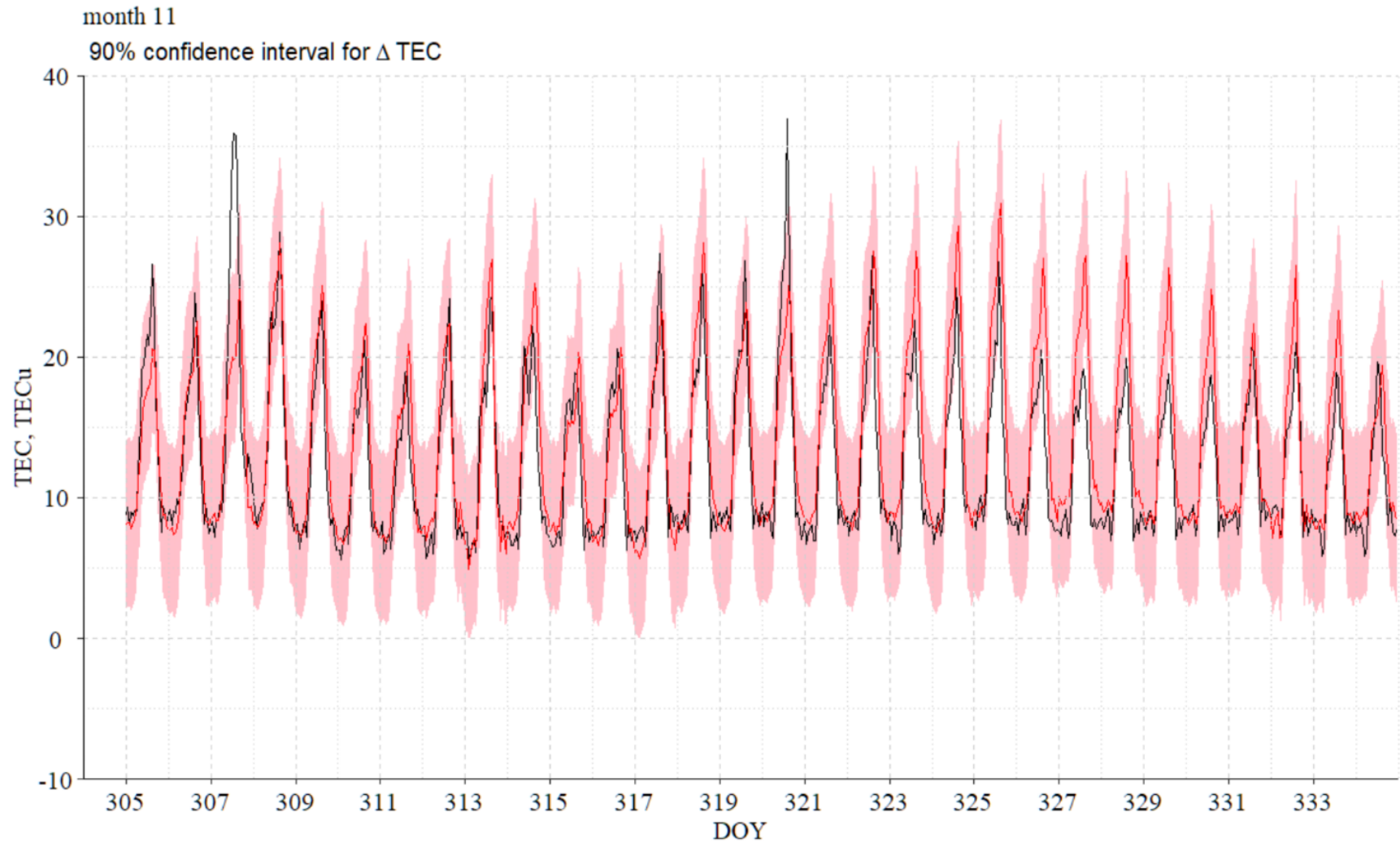

**Figure S13.** Observed (black) and forecasted using the PCA-MRM(L31, mean.lag1.2) model (red) TEC 1h series for November 2015; pink area shows 90% confidence interval.

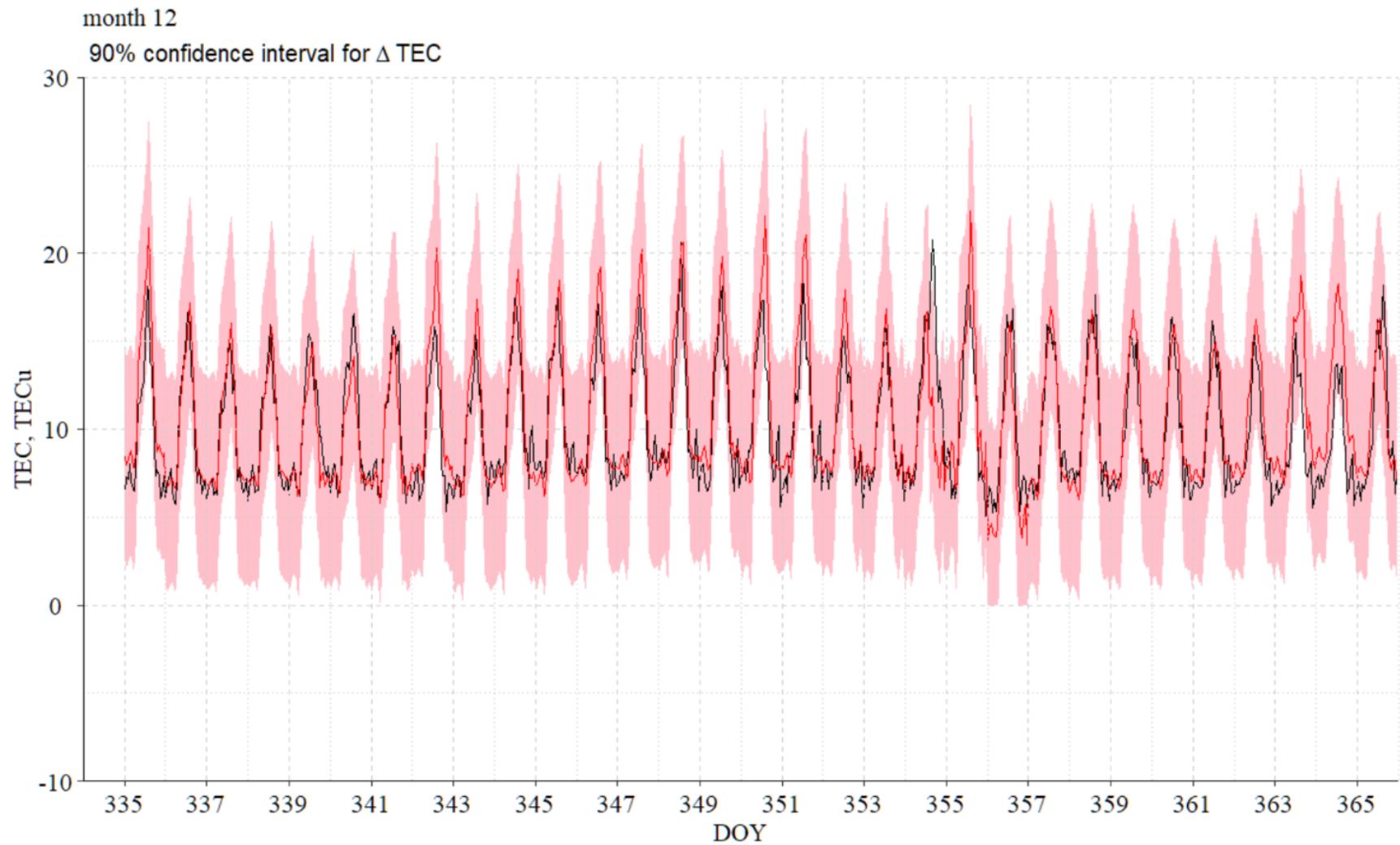

**Figure S14.** Observed (black) and forecasted using the PCA-MRM(L31, mean.lag1.2) model (red) TEC 1h series for December 2015; pink area shows 90% confidence interval.

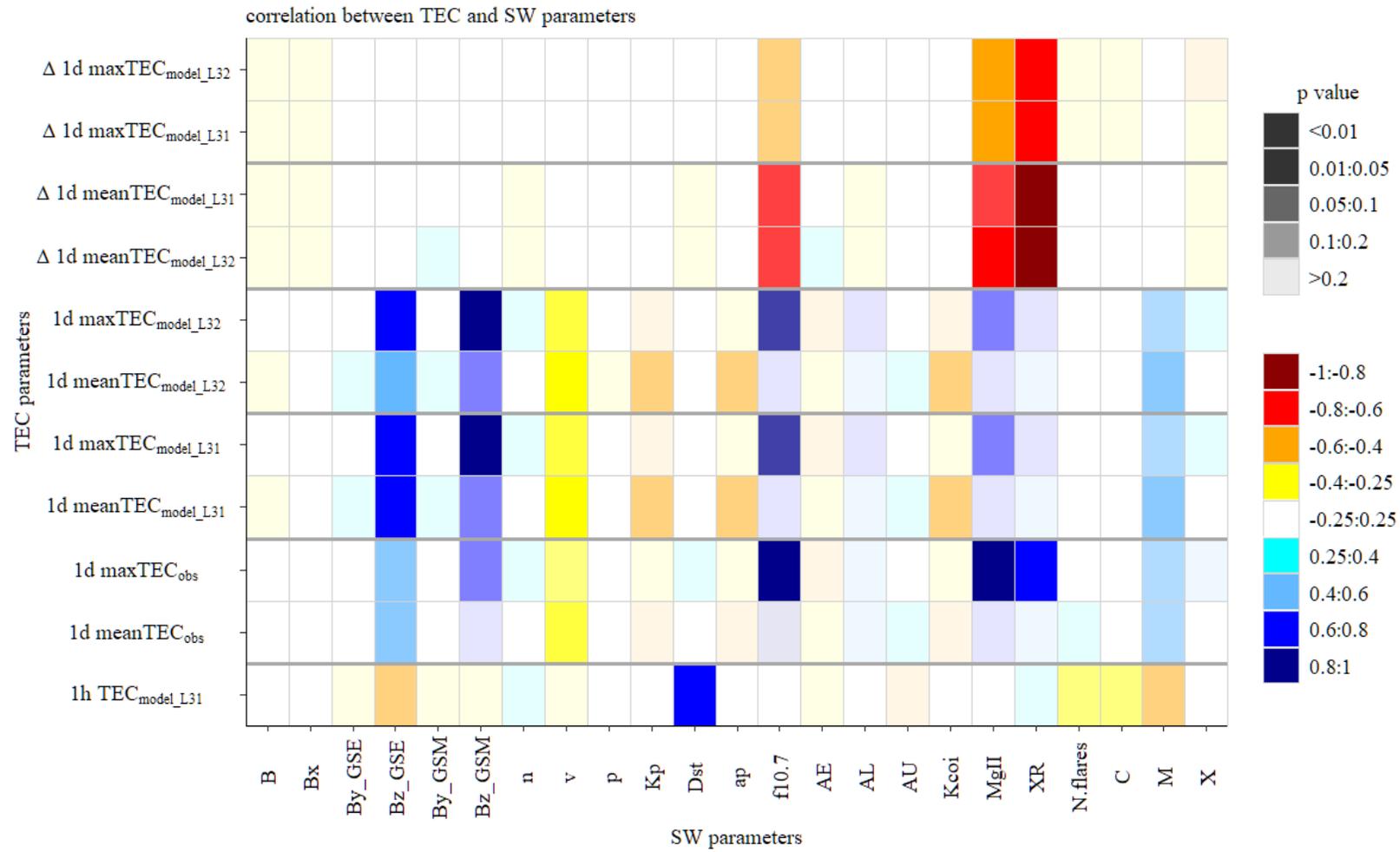

**Figure S15**. Correlation coefficients between TEC parameters: observed and forecasted TEC 1h, 1d mean and 1d max series, and 1d mean and 1d max ΔTEC series (Y-axis) vs SWp (X-axis) on the monthly time scale: r are shown as colours and p values are shown as colour intensity. PCA-MRM models are PCA-MRM(L31, mean.lag1.2) and PCA-MRM(L32, mean.lag1.2).